\documentclass[linenumbers]{aastex631}

\shorttitle{GeV from the supernova remnant G51.26+0.11}
\shortauthors{Araya}

\graphicspath{{./}{figures/}}

\usepackage{tikz,xcolor,hyperref}
\usepackage{subfigure}

\begin{document}

   \title{GeV emission in the region of the supernova remnant G51.26+0.11}

   

\author[0000-0002-0595-9267]{M. Araya}
\affiliation{Escuela de F\'isica, Universidad de Costa Rica \\
San Pedro de Montes de Oca, 11501-2060 San Jos\'e, Costa Rica}


%



\begin{abstract}

The supernova remnant (SNR) G51.26+0.11 was recently discovered and little is known about its properties and environment. Using data from the \emph{Fermi} Large Area Telescope we study the GeV emission seen in the direction of G51.26+0.11 to constrain the origin of the gamma rays and their possible relation to this SNR or to a star-forming region. We also search for emission from molecular gas in the region that could provide dense material for the production of gamma rays. By modeling the multi-wavelength spectrum of G51.26+0.11 from radio to gamma rays we derive the properties of the particle populations that could produce the emission in several possible scenarios. We rule out the star-forming regions (such as G051.010+00.060) seen nearby as the origin of the GeV emission. The correspondence seen between the gamma-ray and radio morphologies support a scenario where G51.26+0.11 is the cause of the gamma rays. The flat spectral energy distribution observed at GeV energies is best fit by hadronic or inverse Compton emission, while a bremsstrahlung model cannot properly account for the radio fluxes under a simple one-zone scenario. A pulsar wind nebula origin of the high-energy photons cannot be ruled out or confirmed.


\end{abstract}


\section{Introduction}
Supernova remnants (SNRs) play an important role in the Galaxy. They heat up the interstellar medium (ISM), influence star formation, distribute heavy elements and accelerate cosmic rays. Given the supernova rate expected and seen in the Galaxy \citep[e.g.,][]{1994ApJ...425..205V} several thousand SNRs should be observed and yet only a few hundred are known \citep{2019JApA...40...36G}. Selection effects such as the lack of sensitive radio continuum observations are thought to be the cause of the discrepancy. New candidate SNRs have been discovered recently with the increasing number of observations across the electromagnetic spectrum, for example The HI, OH, Recombination line survey of the Milky Way \citep[THOR,][]{2017A&A...605A..58A_THOR}, has found 76 candidate SNRs.

SNRs accelerate particles to relativistic energies, which can then be detected through their gamma radiation. At high (GeV) and very high (TeV) energies source catalogs by the \emph{Fermi} Large Area Telescope \citep[LAT,][]{2016ApJS..224....8A,2020ApJS..247...33A}, the High Energy Stereoscopic System \citep{2018A&A...612A...1H} and the High-Altitude Water Cherenkov Observatory \citep{2020ApJ...905...76A} reveal SNRs as well as several unassociated sources which could be previously unknown SNRs. The study of the gamma-ray emission from SNRs is important to understand the properties of the particle populations that they accelerate, and multi-wavelength observations are useful to constrain SNR parameters such as distance and ambient density, which are usually poorly known.

G51.21+0.11 is one of the SNR candidates discovered by THOR through combined mid-infrared and radio observations \citep{2017A&A...605A..58A_THOR}. The radio continuum observations originally revealed a region with an extension of $\sim 30\arcmin$ and a flux density of $24.35\pm 2.10$ Jy at a frequency of 1.4 GHz. Follow up radio observations by \cite{2018ApJ...860..133D} revealed a complex morphology and a radio spectral index $\alpha = -0.7 \pm 0.21$ for G51.21+0.11 (the radio flux density $S_\nu \propto \nu^{\alpha}$ where $\nu$ is the frequency), making it a good SNR candidate. They were not able to confirm its nature due to the lack of data at other wavelengths that are relevant for SNR studies, particularly in the X-ray band.

Using radio observations at lower frequencies, \cite{2018A&A...616A..98S} claimed that a small southern portion of G51.21+0.11 is in fact a different SNR, which they labeled G51.04+0.07. They found that G51.04+0.07 is located at a distance of $7.7\pm 2.3$ kpc, and it is also likely closer than the most prominent HII region seen near in the sky, known as G051.010+00.060 \citep{Anderson_2014,2019ApJS..240...14L}, which is a source of thermal radio continuum emission. At the same time, \cite{2018ApJ...866...61D} confirmed the presence of nonthermal radio emission from G51.04+0.07 and also from part of the source to the north of it at the location of the originally proposed SNR candidate G51.21+0.11. Given the identification of the source G51.04+0.07 as an independent SNR, the rest of the emission was reported as a separate shell-type SNR, which \cite{2018ApJ...866...61D} labeled G51.26+0.11. This SNR has a proposed radius of $11.3\arcmin$ \citep{2018ApJ...866...61D}. Recently, using data from the Karl G. Jansky Very Large Array GLObal view of the STAR formation in the Milky Way (GLOSTAR), \cite{2021arXiv210306267D} measured the degree of polarization and radio spectral indices of G51.26+0.11 and G51.04+0.07, confirming their nature as SNRs.

The First Fermi LAT Supernova Remnant Catalog \citep{2016ApJS..224....8A} did not explore the gamma-ray emission in the region of G51.21+0.11 as it predates its discovery. The \textit{Fermi} Large Area Telescope Fourth Source Catalog Data Release 2 \citep[4FGL-DR2,][]{2020arXiv200511208B} shows two point sources having no association in the region of G51.26+0.11, which are 4FGL J1925.4+1616 and 4FGL J1924.3+1628. The brightest of the two sources, 4FGL J1925.4+1616, has been a target of a 5.2 ks-\emph{Swift} X-Ray Telescope observation as part of a survey of \emph{Fermi} unassociated sources \citep{2013ApJS..207...28S}. In a follow-up study of these sources carried out by \cite{2019ApJ...887...18K} a signal-to-noise threshold of 4 is defined as a lower limit to consider a source as detected, a criterion that is not satisfied by 4FGL J1925.4+1616\footnote{See https://www.swift.psu.edu/unassociated/}.

In this paper we carry out a detailed study of the gamma-ray emission detected by \emph{Fermi} in this region to obtain its morphology and spectrum. The possible connection of the gamma rays to any SNR, known or otherwise, the star-forming region G051.010+00.060 or the pulsars seen in the vicinity is probed. We compare the locations of the gamma rays with that of the radio emission by the THOR survey \citep{2016A&A...595A..32B}. Both star-forming regions and SNRs are known to produce gamma rays through inelastic collisions between relativistic protons or nuclei and ambient matter. Dense molecular clouds in the vicinity of cosmic ray accelerators thus provide targets for this hadronic interactions. We explored data from the $^{13}$CO (J = $1\to 0$) line emission of the Boston University-Five College Radio Astronomy Observatory Galactic Ring Survey \citep[GRS,][]{2006ApJS..163..145J} to estimate the sizes and densities of the molecular clouds in the region of G51.26+0.11 and G051.010+00.060. We also compare these observations with $^{12}$CO (J = $1\to 0$) data from \cite{2001ApJ...547..792D}. Finally, we modeled the multi-wavelength spectrum using hadronic and leptonic scenarios to derive the parameters of the relativistic particles responsible for the emission under the one-zone approximation to try to constrain its origin.

\section{\emph{Fermi}-LAT observations and data analysis}\label{sec:lat}
The LAT onboard the \emph{Fermi} satellite continually scans the sky and detects photons in the energy range from about 20 MeV to more than 300 GeV. We used data collected between the beginning of the mission, August 2008, to December 2020 (i.e., with an additional 2.3 yr of data compared to the 10-yr 4FGL-DR2 catalog). We included events in the energy range from 200 MeV to 500 GeV reconstructed within $15\degr$ of the center of our region of interest (ROI), located at the coordinates RA$=291.3\degr$, Dec$=16.3\degr$ (J2000). This is approximately the reported position of the source 4FGL J1925.4+1616. We used {\tt fermitools} version 2.0.0, a publicly available software for treating \textit{Fermi} data, and the open-source {\tt PYTHON} package {\tt fermipy} version 1.0.0, to perform the analysis. We applied the recommended quality cuts for {\tt PASS8} data analysis, including a zenith angle cut of $90\degr$ to avoid gamma-ray contamination from Earth's limb, {\tt DATA\_QUAL>0} and {\tt LAT\_CONFIG==1}. The analysis used the {\tt P8R3\_SOURCE\_V3} instrument response functions and we combined back and front-converted events in the {\tt SOURCE} class (using the parameters {\tt evtype=3} and {\tt evclass=128}). We adopted the binned maximum likelihood method \citep{1996ApJ...461..396M} to derive the spectral and morphological parameters of the sources. The significance of a new source with one additional parameter with respect to the model without the source (known as the null hypothesis) can be estimated with the square root of the test statistic, TS $=-2\times \ln (\mathcal{L}_0/\mathcal{L})$, where $\mathcal{L}_0$ and $\mathcal{L}$ are the values of the maximum likelihoods for the null hypothesis and for a model with the additional source, respectively.

To model the background we included the sources in the ROI listed in the 4FGL-DR2 catalog along with the recommended Galactic and isotropic diffuse emission components, described by the files {\tt gll\_iem\_v07.fits} and {\tt iso\_P8R3\_SOURCE\_V3\_v1.txt}, respectively, and provided with the {\tt fermitools}. The energy dispersion correction was applied as recommended by the LAT team\footnote{See https://fermi.gsfc.nasa.gov/ssc/data/analysis/documentation/Pass8\_edisp\_usage.html}. The sources 4FGL J1925.4+1616 and 4FGL J1924.3+1628 were not included in the models in order to study the GeV emission in the region more carefully.

\subsection{Morphological analysis}
\begin{figure*}
   \centering
     \subfigure[]{\includegraphics[width=0.8\textwidth]{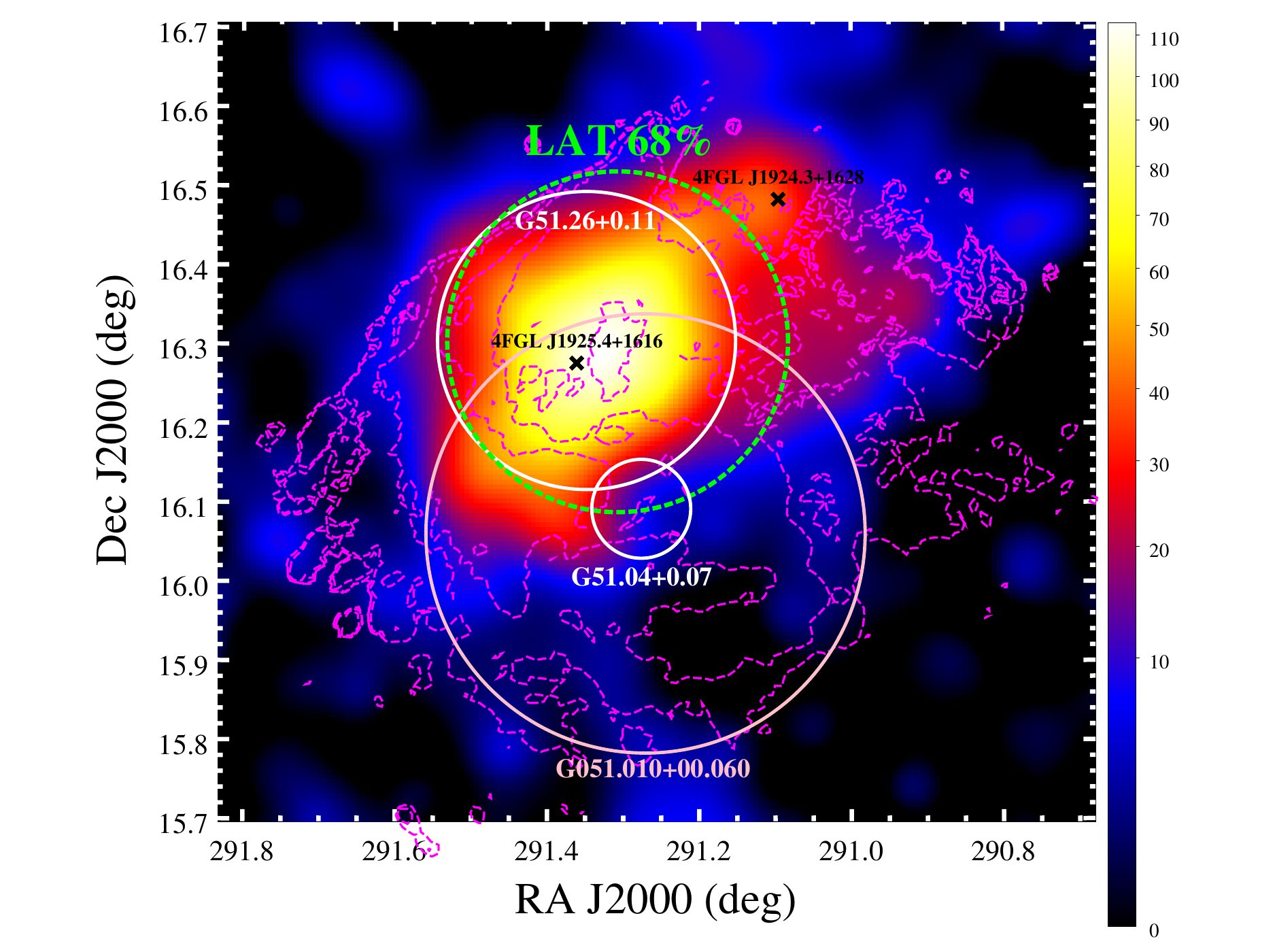}}
     \subfigure[]{\includegraphics[width=0.8\textwidth]{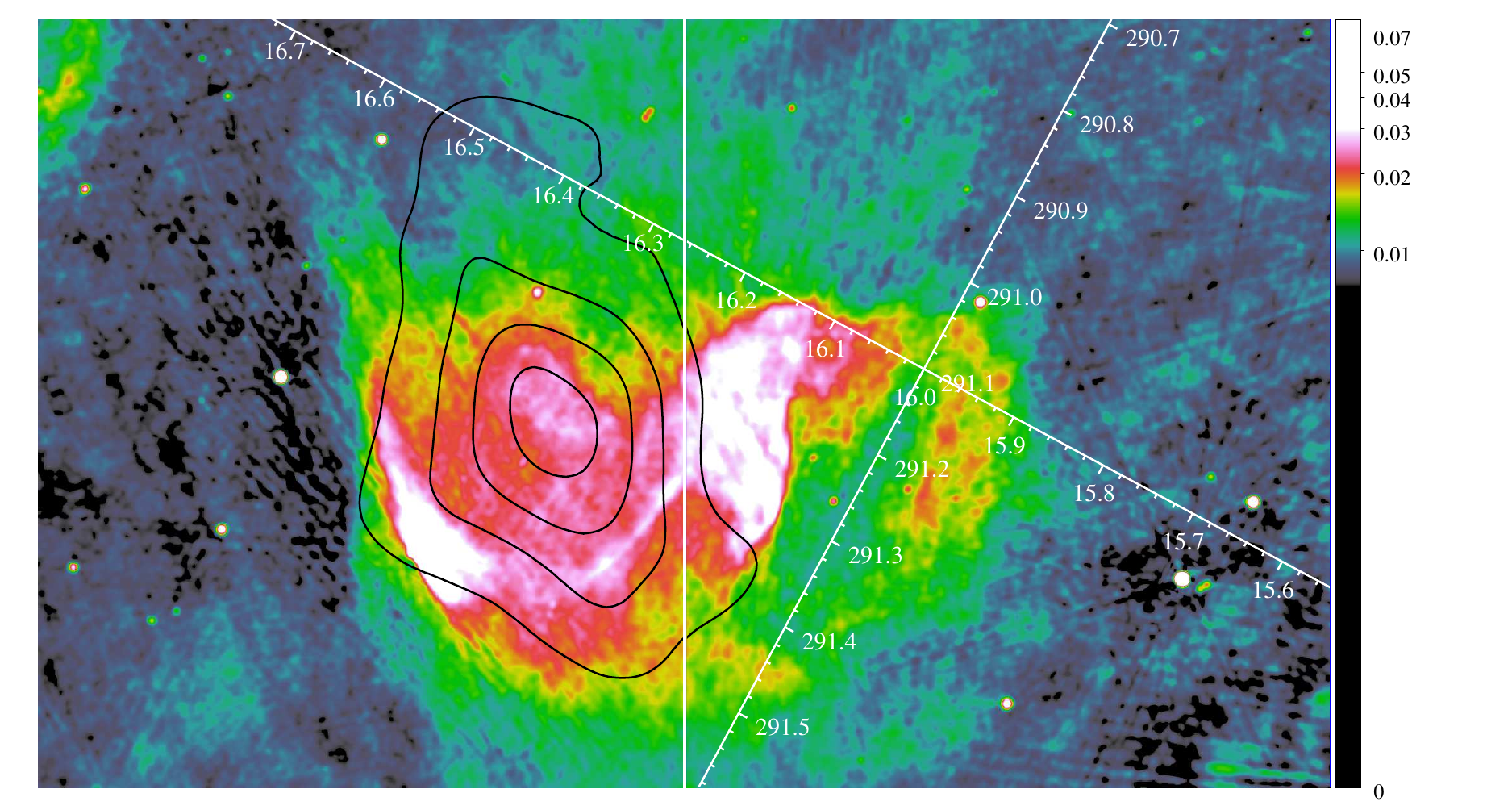}}
\caption{(a) LAT TS map for a point source hypothesis with events having energies above 5 GeV showing the excess emission above the background. The locations and sizes of G51.26+0.11 and G51.04+0.07, according to \cite{2018ApJ...866...61D}, are shown, as well as those of the star-forming region G051.010+00.060. The dashed circle represents the 68\%-containment region of the Gaussian source found in this work to model the GeV emission. The positions of the two 4FGL sources originally found in the LAT catalogs are also shown for comparison. The contours represent the radio continuum emission from THOR observations \citep{2016A&A...595A..32B} in logarithmic scale from 0.01 to 0.085 Jy/beam, possibly a mixture of thermal and non thermal emission. The color (square root) scale is in units of TS. (b) The radio continuum emission from THOR with the TS contours from the gamma-ray TS map above at the levels of 25, 50, 75 and 100 overlaid. The (logarithmic) color scale is in Jy/beam and the axes coordinates are in degrees.}\label{fig1:tsmap}
    \end{figure*}

In order to take advantage of the improved spatial resolution of the LAT at higher energies, we performed a study of the morphology of the gamma-ray emission in the region of G51.26+0.11 using events with energies above 5 GeV. We first searched for an appropriate model describing the background sources. As a first step, we fit the spectral normalizations of the sources located within $10\degr$ of the ROI center simultaneously as well as all the spectral parameters of the sources found within $3\degr$ of the ROI center. As a second step, we searched for additional local TS maxima likely associated to new uncatalogued sources in the ROI with the {\tt find\_sources} routine of {\tt fermipy}. We incorporated the newly found sources into the background model and proceeded to fit their spectral parameters, together with the free parameters of the other known sources. In order to visualize the residual gamma-ray emission for the source of interest in the center of the ROI, a TS map of the null hypothesis was constructed by fitting the normalization of a test point source having a power law spectral model and a fixed index of 2, whose position is moved through each pixel in the map. The result is seen in Fig. \ref{fig1:tsmap}. The first image shows the locations and sizes of the SNRs G51.26+0.11 and G51.04+0.07, the star-forming region G051.010+00.060 and the GeV source found in this section to represent the gamma-ray emission. The contours show the radio continuum emission from the THOR survey \citep{2016A&A...595A..32B}. The gamma-ray emission has a maximum significance at the location of the SNR. In the second image, the radio continuum emission from THOR is shown in comparison with the gamma-ray TS contours.

Different morphological models were fitted to the GeV emission to choose the best available representation. In order to do this, the Akaike Information Criterion \citep[AIC,][]{1974ITAC...19..716A} was calculated for each case. This is defined as AIC = $2k - 2\ln(\mathcal{L})$, where $k$ is the number of free parameters in the model and $\mathcal{L}$ the maximum likelihood obtained in the fit. The definition of this statistical criterion is such that the best available model is the one that minimizes the AIC. The models used were two point sources, a Gaussian and a disk, for which the extensions and locations were optimized to maximize the likelihood. The locations of the two point sources were obtained by re-optimizing the positions of the catalogued sources 4FGL J1925.4+1616 and 4FGL J1924.3+1628. The morphological parameters were optimized together with the spectral parameters of the sources, assumed to be described by simple power law spectra (this choice of spectral function is justified below) given by $$\frac{dN}{dE} = N_0 \left( \frac{E}{E_0} \right)^{-\Gamma},$$ with $E_0 = 5000$ MeV a fixed scale factor.

Table \ref{tab1:AIC} shows the results for each morphological hypothesis representing the GeV emission in the region of G51.26+0.11. As can be seen, the Gaussian model offers a better description of the data compared to the other models and it is thus chosen for the rest of the analysis. The sizes, locations and best-fit spectral parameters for each model are shown for reference. The centroids of the disk and Gaussian are consistent with the optimized location of the source 4FGL J1925.4+1616, which is the brightest of the two original 4FGL sources. The newly optimized positions of these two sources are also consistent within the statistical uncertainties with their positions reported in the 4FGL-DR2 catalog. The spectral index found here above 5 GeV for 4FGL J1924.3+1628 ($1.90\pm0.05$) is consistent with the cataloged value ($1.7\pm 0.2$), while that for 4FGL J1925.4+1616 ($2.08\pm0.03$) is lower compared to the cataloged value ($2.36\pm 0.07$). This could be due to the different energy ranges used in the present morphological study (5--500 GeV) compared to the wider interval used for the catalog (100 MeV -- 1 TeV). The measured flux levels obtained here (quantified by the normalization of the spectral density distribution, $N_0$, once corrected to account for the different scale factors $E_0$), however, are comparable to the reported 4FGL-DR2 values. The fluxes obtained from the extended templates, as expected, are higher than those of the point sources. We also obtained significance maps of the residual gamma-ray emission resulting from each of the morphological models which show that the Gaussian model more properly accounts for the emission in the region. In order to estimate the significance of the source extension, we follow \cite{2012ApJ...756....5L} and calculate TS{\tiny ext}$=2\ln(\mathcal{L}_{\mbox{\tiny ext}}/\mathcal{L}_{\mbox{\tiny ps}})$, where $\mathcal{L}_{\mbox{\tiny ext}}$ and $\mathcal{L}_{\mbox{\tiny ps}}$ are the likelihoods resulting from fitting the extended source and a point source. The resulting value, TS{\tiny ext}$=32.3$, indicates that the extension is significant beyond a point source for the LAT.

We note that even though with the current resolution provided by the LAT an extended model is preferred over two point sources, future observations with better resolution might reveal an extended source at the location of 4FGL J1925.4+1616 and an unrelated point source at the location of 4FGL J1924.3+1628.

\begin{table}
\caption{\label{tab1:AIC}Results of the morphological analysis of the GeV emission in the region of G51.26+0.11.}
\centering
\begin{tabular}{lcccc}
\hline\hline
Model                 & RA, Dec ($\degr$) & Extension ($\degr$) & Spectral parameters & $\Delta$AIC \\
\hline
Two point sources     &                    &  &                     &   13.8\\
4FGL J1925.4+1616     & $291.32\pm0.01$, $16.29\pm0.01$   &--&  $N_0 = 0.52\pm0.02$, $\Gamma=2.08\pm0.03$  & \\
4FGL J1924.3+1628     & $291.12\pm0.03$, $16.47\pm0.02$   &--&  $N_0 = 0.16\pm0.01$, $\Gamma=1.90\pm0.05$  & \\
\hline
Disk                  &$291.29\pm 0.02\degr$, $16.29\pm 0.02 \degr$ &$0.22^{+0.02}_{-0.01}\,\degr$ & $N_0=1.12\pm0.04$  &  6.6\\
                      &                                              &                              &$\Gamma = 2.12\pm 0.02$ & \\
\hline
Gaussian              &$291.31\pm 0.02\degr$, $16.30\pm 0.02 \degr$  &$0.22^{+0.04}_{-0.03}\,\degr$ & $N_0=1.14\pm0.04$   & 0  \\
                      &                                              &                              & $\Gamma = 2.12\pm0.02$ & \\
\hline
\end{tabular}
\tablecomments{The values of $N_0$ are given in units of $10^{-13}$ MeV$^{-1}$cm$^{-2}$s$^{-1}$. The extension for the disk and Gaussian refers to their respective 68\%-containment radii. $\Delta$AIC is defined as the difference between the AIC value for the given model and that of the model with the lowest AIC value. $1\sigma$ statistical uncertainties are given for morphological and spectral parameters. All parameters in this table were determined in the analysis of events with energies above 5 GeV.}
\end{table}

As shown in Table \ref{tab1:AIC}, the best-fit coordinates of the Gaussian centroid and 68\%-containment radius (with their $1\sigma$ statistical uncertainties) are RA=$291.31\pm 0.02\degr$, Dec=$16.30\pm 0.02 \degr$ (J2000) and $0.22^{+0.04}_{-0.03}\,\degr$. As mentioned earlier, the 68\%-containment region of the best-fit Gaussian is also shown in Fig. \ref{fig1:tsmap}.

\subsection{Spectral analysis}
Once an appropriate model for the morphology of the GeV emission is obtained, we apply this model to events in the ROI with an energy above 200 MeV and obtain the spectrum of the source. The Gaussian template found in this section was added to the model, and a search for new point sources in the ROI was carried out to improve the background description. Once the new point sources are added, we let the normalizations of the sources located within $10\degr$ of the ROI center and the other spectral parameters of the sources located within $5\degr$ of the ROI center free to vary.

Two phenomenological spectral energy functions were used in separate fits to the emission seen in the region of G51.26+0.11 in order to compare the results, a simple power law and a log parabola. Since these two spectral shapes correspond to nested models, comparing the likelihood values obtained in the fits allows to quantify the significance of curvature (i.e., deviation from a power law) in the spectrum. It was found that the addition of a parameter using the log parabola produces a negligible change ($\sim0.2$) in the log-likelihood function, indicating that the spectrum of the source is not significantly curved in the LAT energy range. This justifies the use of a power law for the spectrum in the morphological analysis using events with energies above 5 GeV. If the spectral function is a power-law as defined earlier, with $E_0 = 5000$ MeV a fixed scale, the resulting best-fit values obtained are $N_0 = (1.21 \pm 0.07_{\mbox{\tiny stat}} \pm 0.50_{\mbox{\tiny sys}})\times 10^{-13}$ MeV$^{-1}$ cm$^{-2}$ s$^{-1}$ and $\Gamma = 2.18 \pm 0.04_{\mbox{\tiny stat}} \pm 0.13_{\mbox{\tiny sys}}$. The overall test statistic value of the source is TS$=443$ above 200 MeV, which translates to a detection significance of $21\sigma$. The residual map obtained when adding the source shows no significant emission leftover in the region, meaning that the model found for the source is satisfactory. A spectral energy distribution (SED) was obtained by measuring the flux of the source in ten logarithmically-spaced energy bins. In each bin the normalization of the source of interest was fit together with the normalizations of the sources located within $2\degr$ of the center of the ROI, as well as those of the diffuse and isotropic backgrounds. The spectral index of the source was kept fixed to 2, but the results were not significantly affected by using a different value for the spectral index. If the TS of the source in a bin was below 4, a 95\% confidence level upper limit on the flux was estimated for that bin.

Two factors were considered for calculating the systematic errors in the spectral parameters. The effect of the uncertainty in the Galactic diffuse emission model, which is particularly important for extended sources, was estimated with the use of the eight alternative models developed by \cite{2016ApJS..224....8A} in making the LAT catalog of SNRs. The files were scaled appropriately to account for differences in energy dispersion between Pass 7 and Pass 8 reprocessed data\footnote{See https://fermi.gsfc.nasa.gov/ssc/data/access/lat/Model\_details/Pass8\_rescaled\_model.html}. The uncertainties were calculated as in \cite{2016ApJS..224....8A} for the source parameters in the global fit using the entire energy range as well as in the individual energy bins used for obtaining the SED. The uncertainties in the effective area of the LAT were also propagated onto the spectral parameters of the global fit, as well as to the normalizations of the SED fluxes, using a set of bracketing response functions as recommended by \cite{2012ApJS..203....4A}. In these alternative fits the pivot energy was used as the value of the scale parameter $E_0$, which was estimated with the covariance error matrix of the global fit. For the individual SED fluxes the statistical and systematic errors were combined in quadrature. Finally, we also estimated the effect of the Galactic diffuse emission uncertainties in the source extension. In a manner analogous to the treatment of uncertainties for spectral parameters explained before, we fit the extension of the Gaussian template using events above 5 GeV for the alternative background models and calculated a systematic error on the 68\%-containment radius of $0.10\degr$. The source was found to be significantly extended in all cases.

\section{Properties of the molecular gas}\label{sec:ism}
\begin{figure*}
     \centering
     \includegraphics[width=0.95\textwidth]{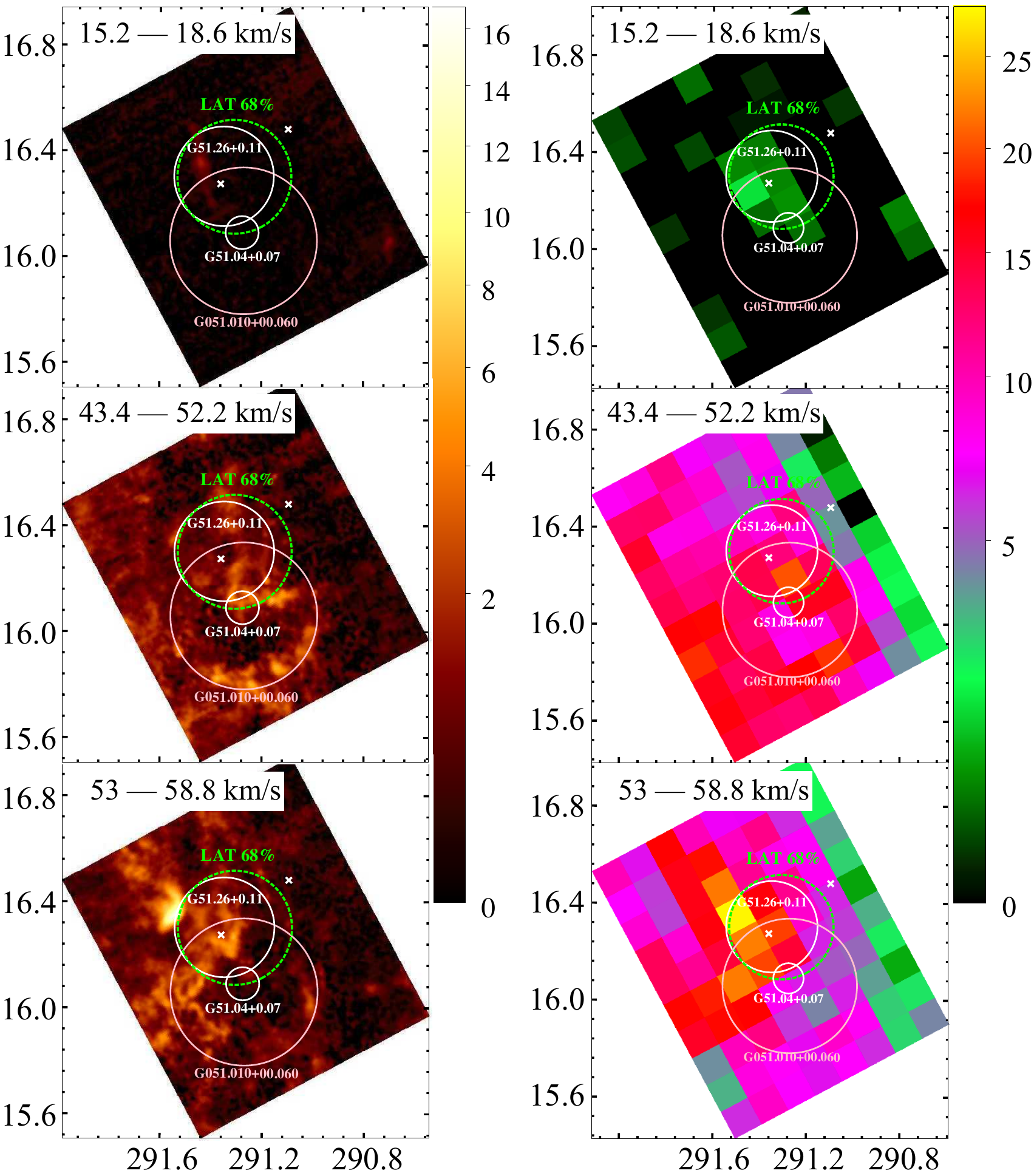}
     \caption{Integrated CO emission in the velocity intervals where gas is seen in projection within the 68\%-containment size of the GeV source (shown in the images). $^{13}$CO (J = $1\to 0$) data from the GRS survey \citep{2006ApJS..163..145J} is shown in the left panels whereas the $^{12}$CO (J = $1\to 0$) observations from \cite{2001ApJ...547..792D} are shown in the right panels. The regions are the same as in Fig. \ref{fig1:tsmap}, with the crosses representing the positions of the original 4FGL sources. The color (square root) scale is in units of K km s$^{-1}$ in both sets of maps. The vertical and horizontal axes are the declination and right ascension (J2000), respectively, in degrees.}
        \label{fig2:CO}
\end{figure*}

The average brightness temperature from the $^{13}$CO (J = $1\to 0$) emission in the region within the 68\%-containment radius of the LAT source shows three distinctive peaks in the velocity intervals 15.2--18.6 km s$^{-1}$, 43.4--52.2 km s$^{-1}$ and 53--58.8 km s$^{-1}$ (the line velocities, $v_{lsr}$, used in this work are all measured with respect to the local standard of rest). The same conclusion is obtained after analyzing the observations of the $^{12}$CO (J = $1\to 0$) line emission. Although the spatial resolutions of both surveys are very different, maps comparing the integrated CO line intensities from both data sets are shown in Fig. \ref{fig2:CO}.

Adopting a standard Galactic rotation model \citep{1993A&A...275...67B} we calculated the possible kinematic distances associated to the midpoints of the velocity intervals for the Galactic coordinates $l=51^\circ$, $b=0.1^\circ$. According to the rotation model and using a distance of $R_0=8.15$ kpc from the Sun to the Galactic center and a circular rotation speed at the position of the Sun of $\Theta_0=236$ km s$^{-1}$ \citep{2019ApJ...885..131R} the maximum predicted velocity for these coordinates is $\sim 54$ km s$^{-1}$. The highest integrated intensity for gas seen at the location of the gamma-ray source is seen in the data for the velocity interval 53--58.8 km s$^{-1}$, indicating a slight discrepancy with the Galactic rotation model adopted. However, the rotation model of the Galaxy is known to have issues at the location of spiral arms, and estimations of kinematic distances are also affected by turbulent and proper gas motions, which could account for the discrepancy. We only limit ourselves to cite the possible kinematic distances predicted by the standard rotation model and leave a more detailed study of these effects for the future. As will be seen, the conclusions in this work are valid regardless of the actual source distance. We therefore consider that the gas seen in the velocity interval 53--58.8 km s$^{-1}$ has a kinematic distance of 5 kpc, which is the solution obtained from the rotation model for the maximum allowed theoretical velocity. On the other hand, the near and far kinematic distances of the gas associated to the other velocity intervals in Fig. \ref{fig2:CO} are 1 kpc and 9.2 kpc (taking $v_{lsr} = 17$ km s$^{-1}$ as a representative velocity for this gas) and 3.7 kpc and 6.6 kpc (for $v_{lsr} = 48$ km s$^{-1}$). As seen in Fig. \ref{fig2:CO}, there is a correspondence between the molecular gas in the line velocity interval $v_{lsr} = 43.4-52.2$ km s$^{-1}$ and the HII region G051.010+00.060. This star-forming region is located at a distance of $7.2\pm1.2$ kpc \citep{2014ApJS..212....1A_wise_catalog}, which is compatible with the far kinematic distance determined here for the gas.

The GRS observations have an adequate resolution \citep[$46\arcsec$,][]{2006ApJS..163..145J} to estimate the size and average integrated intensity of the clouds of gas seen within the 68\%-containment radius of the GeV source \citep[note that this region has a larger angular size than the assumed size of the SNR shell, taken to be $11.3\arcmin$,][]{2018ApJ...866...61D}. From the GRS data we calculated the average integrated intensity of the $^{13}$CO (J = $1\to 0$) transition line (labeled $W_{^{13}CO}$), and the appropriate gas filling factor using the actual area covered by the clouds, which are then modeled as spheres. This is done for each velocity interval, which results in several possible cloud and SNR radii. The molecular hydrogen column density in cm$^{-2}$ is then found with the relation $N_{H_2} = X_{{^{13}CO(1\to 0)}} W_{^{13}CO}$. We adopt the conversion factor $X_{{^{13}CO(1\to 0)}} = 4.92\times 10^{20}$ (K km s$^{-1}$)$^{-1}$ from \cite{2001ApJ...551..747S}. Finally, the mass of gas, $M$, is determined by $$M = 2m_H N_{H_2}A,$$ where $A$ is the cross-sectional area of the cloud and $m_H$ is the mass of a hydrogen atom. The average number density ($n$) can then be estimated by dividing the total mass by the volume of the corresponding cloud and the mass of a hydrogen atom. The results are shown in Table \ref{tab2:density} for the possible near and far kinematic distances determined in this work, together with the resulting physical radius of the SNR for comparison.


\begin{table}[h!]
\caption{\label{tab2:density} Average hydrogen number density ($n$) estimated from $^{13}$CO (J = $1\to 0$) observations.}
\centering
\begin{tabular}{lcccc}
\hline\hline
Distance (kpc) & Velocity interval (km s$^{-1}$) & $r_{\mbox{\tiny SNR}}\,^a$ (pc) & $r_c\,^a$ (pc) & $n$ (cm$^{-3}$)$\,^b$\\
\hline
1.0  &   $15.2-18.6$      &3.3& 1.2 & 360\\
9.2  &                    &30 & 11 & 39\\
\hline
3.7   &   $43.4-52.2$     & 12 & 13 & 82\\
6.6    &                  &21.6 & 24 & 46\\
\hline
5.0     &   $53-58.8$     &16.4 & 17 & 80\\
\hline
\end{tabular}
\tablecomments{$^a r_{\mbox{\tiny SNR}}$ and $r_c$ are the SNR and cloud radius, respectively, for the corresponding distance. \\ $^b$The uncertainty on $n$ resulting from the statistical uncertainties associated with the conversion factor $X_{{^{13}CO(1\to 0)}}$ is of the order of 30\% \citep{2017MNRAS.464.3757L}.}
\end{table}

\section{Discussion}
In this section we present the physical models that account for the spectrum of the GeV emission and discuss possible sources. We made use of the properties of the dense gas in the ISM at the location of the gamma-ray source, shown in Section \ref{sec:ism}. We fit the GeV data using one-zone leptonic and hadronic scenarios to derive the particle distributions required using the {\tt naima} package \citep{naima}. We used the recently measured radio fluxes of G51.26+0.11 from \cite{2021arXiv210306267D}. The 200 MHz and 1.4 GHz flux densities of G51.26+0.11 are, respectively, $25.8 \pm 3.6$ Jy and $12.4 \pm 0.6$ Jy. Radio fluxes measured in the larger region associated to the original THOR candidate, G51.21+0.11 \citep{2017A&A...605A..58A_THOR,2018ApJ...860..133D}, are also shown for comparison in the spectral plots in this section. After modeling the SED the origin of the emission is discussed.

\subsection{Gamma-ray SED model}
In fitting the broadband non-thermal fluxes detected in the region of G51.26+0.11 we tested two different shapes for the particle distribution in energy, namely a power law with an exponential cutoff and a broken power law with an exponential cutoff. These particle distribution shapes are usually used to model the SEDs of SNRs \citep[see, e.g.,][]{2015A&A...574A.100H,2016A&A...586A.134P,2019A&A...623A..86A}. As seed photon fields for the calculation of the inverse Compton scattering fluxes (IC) we adopt the cosmic microwave background (CMB), a far-infrared (FIR) field and stellar optical and near-infrared (NIR) photons. For the latter two, the densities and temperatures adopted are the same as those estimated at a Galactocentric distance of 8 kpc \citep[FIR: 26 K, 0.35 eV cm$^{-3}$, NIR: 2000 K, 0.7 eV cm$^{-3}$,][]{2011ApJ...727...38S}, which we refer to as the standard field values. Most of the parameters do not change substantially if the field properties are modified. In order to explore the effect of the change in the standard field values on the resulting spectral parameters of the particle distribution and ambient magnetic field, the values of the field temperatures and their densities were changed to find the range that produces a variation of at least one of the fitted parameters by $1\sigma$, the corresponding parameter error obtained in the fit using the standard values. We found that the field temperatures could be multiplied by a factor ranging from 0.5 to 3, for the FIR field, and from 0.45 to 2.5 for the NIR field, and their densities by factors of 0.35 to 2 and 0.7 to 1.5, respectively. We found that parameters such as the magnetic field, cutoff energy and spectral indices of the particle distribution are not particularly sensitive to the photon field parameters, and usually changed by less than 20\% with respect to the values found with the standard photon field description. On the other hand, the break energy was more sensitive and could increase by up to 50\% with respect to our reported value when modifying the photon field parameters in the given ranges. In the hadronic scenario, we used the parametrization of the photon production cross section for proton-proton interactions presented by \cite{2014PhRvD..90l3014K}.

\begin{figure*}
     \centering
     \subfigure[]{\includegraphics[width=0.45\textwidth]{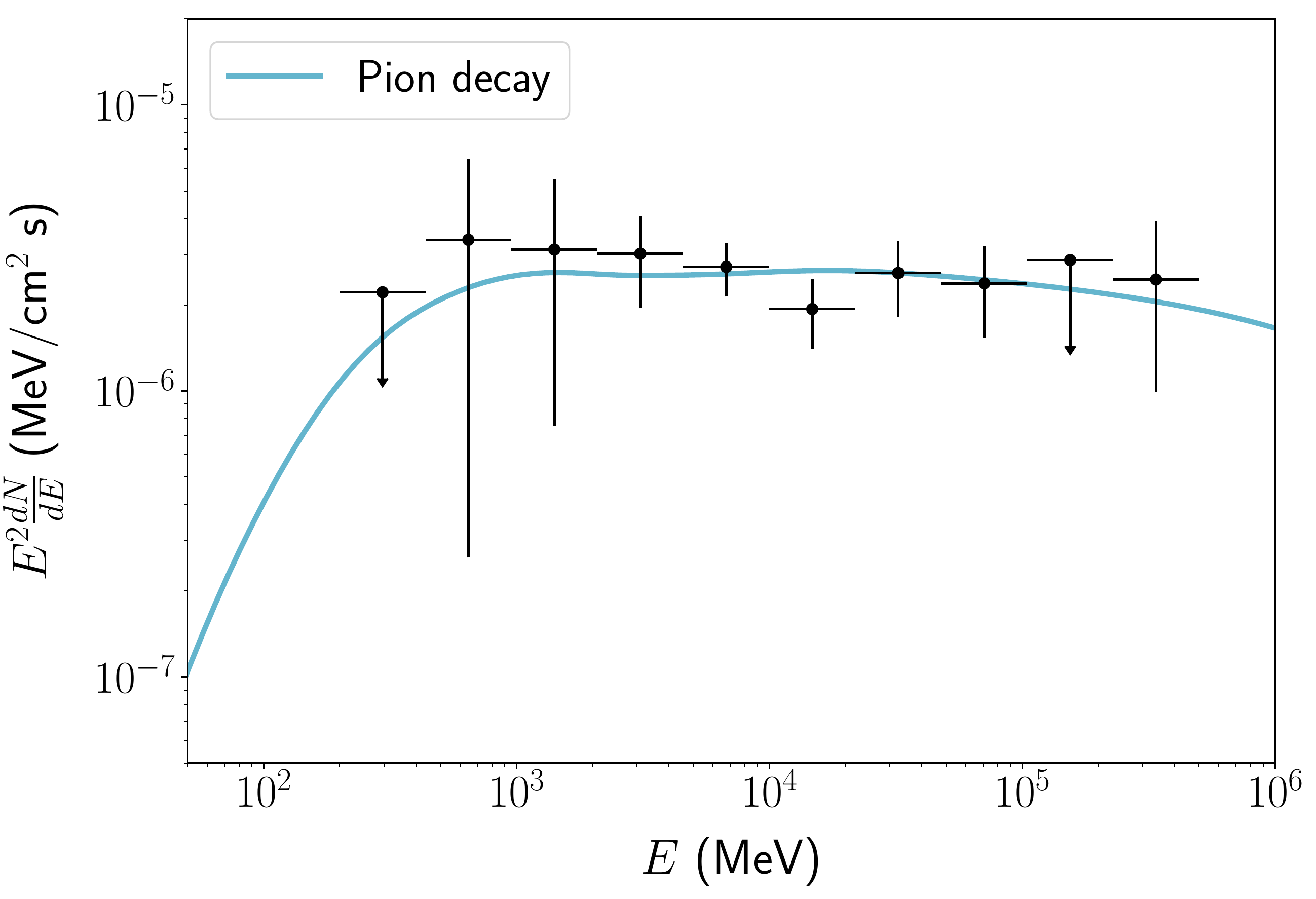}}\\
     \subfigure[]{\includegraphics[width=0.45\textwidth]{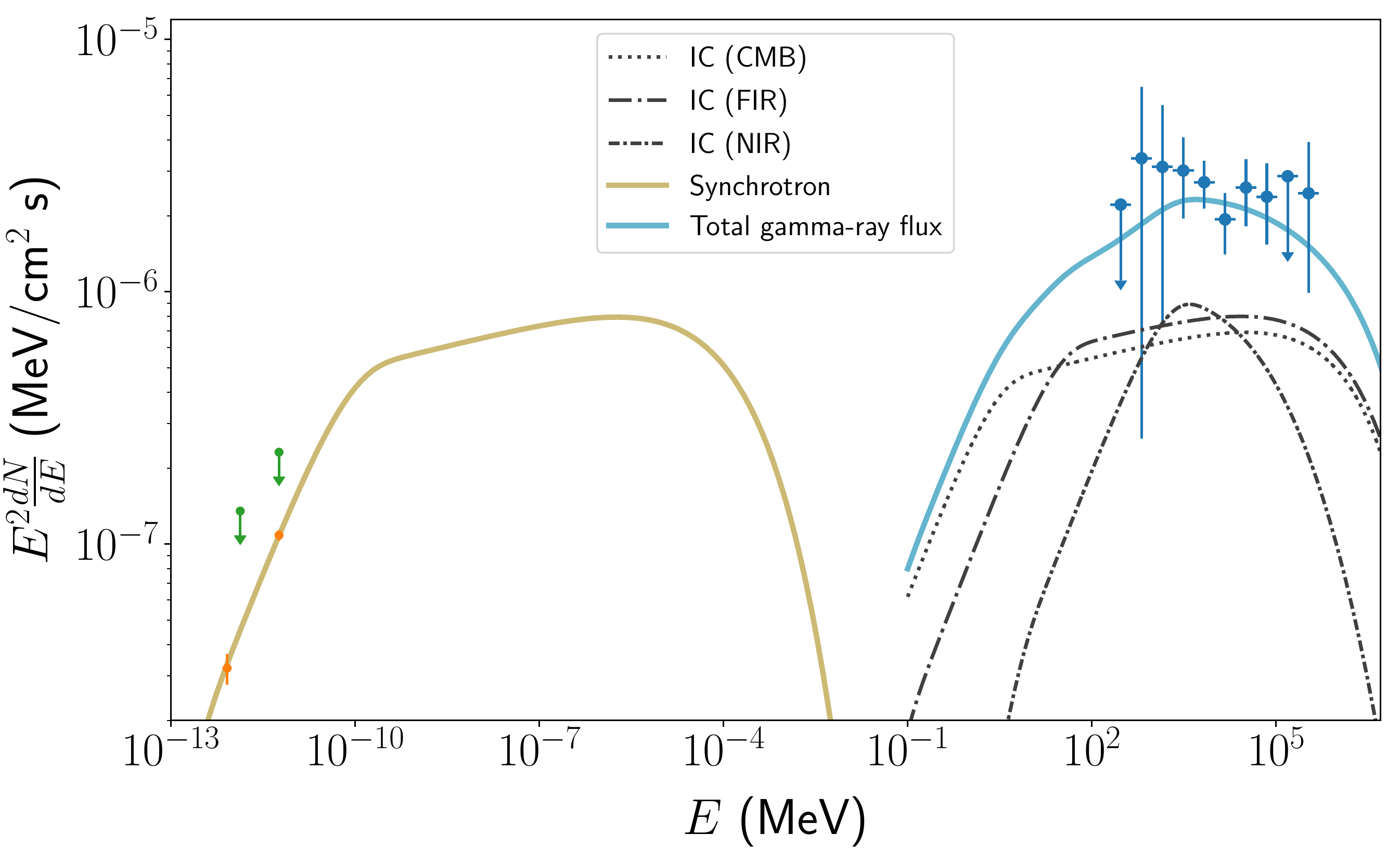}}
     \subfigure[]{\includegraphics[width=0.45\textwidth]{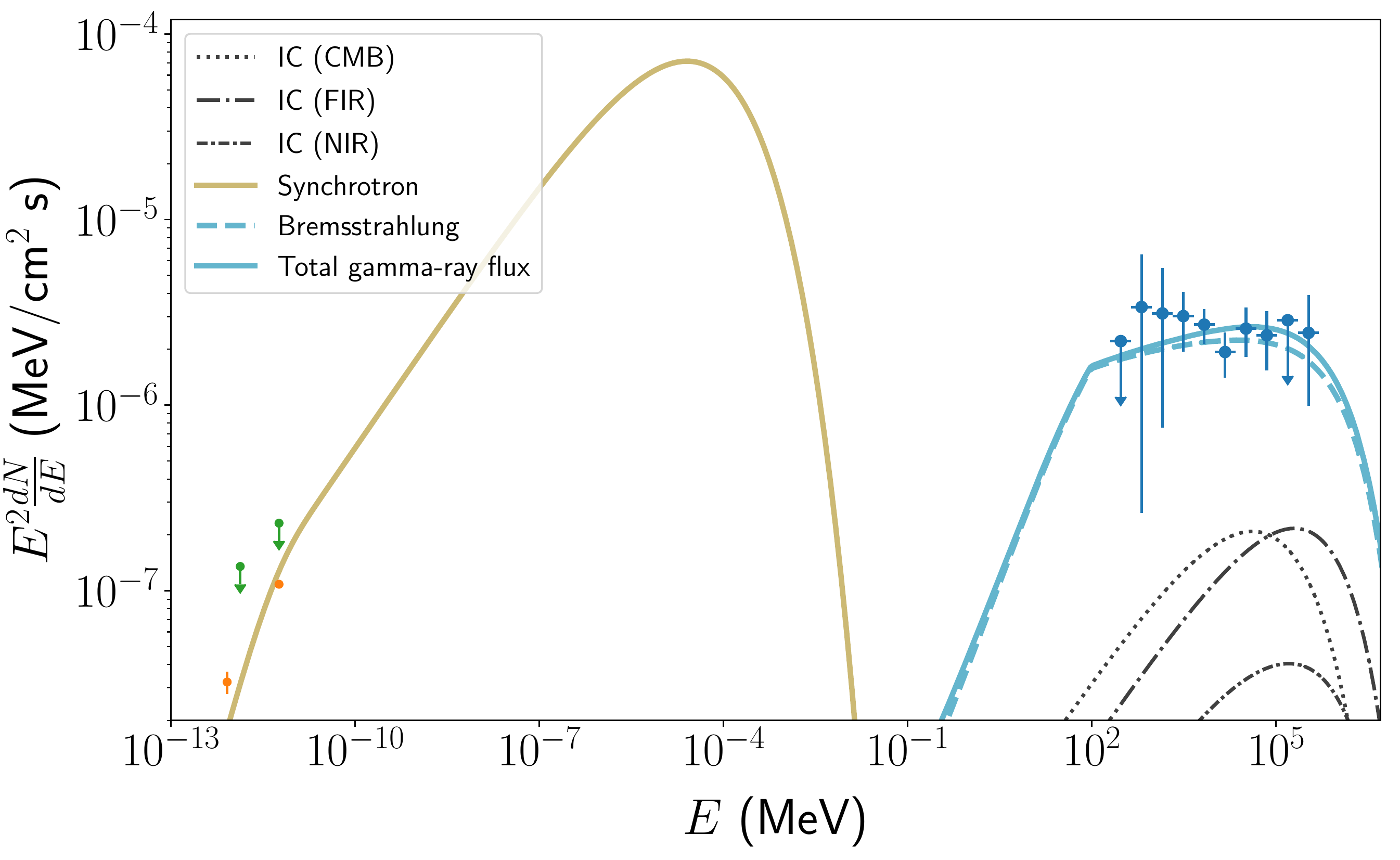}}
     \caption{SED data points and models for each scenario where the high-energy emission results predominantly from neutral pion decay from hadronic interactions by high-energy cosmic rays (a), IC emission from leptons (b) and bremsstrahlung emission from leptons (c). The gamma-ray fluxes were obtained in this work. The radio fluxes used in the fits are from the SNR G51.26+0.11 \citep[measured by][]{2021arXiv210306267D}. The radio fluxes originally reported for the THOR SNR candidate G51.21+0.11 \citep{2017A&A...605A..58A_THOR,2018ApJ...860..133D} are shown as upper limits for comparison.}
        \label{fig4:sed}
\end{figure*}

We perform three independent fits to the gamma-ray SED points obtained in this work under the assumption that one of the three main mechanisms for high energy emission expected to operate in Galactic sources dominates in each case: pion decay emission from hadronic processes, IC from leptons interacting with ambient photons and bremsstrahlung emission also from high energy leptons interacting with ambient material. We note that the overall SED could also be explained by combinations of these processes or from more than one particle population, but given how little is known about the source and its environment our aim is to probe the properties of the particles assuming, as a first approximation, that the most simple scenarios hold. Fig. \ref{fig4:sed} shows the SED and the resulting best-fit models in each case. The AIC calculated from the corresponding likelihood function is used to select the best fits.

In the hadronic scenario for the gamma rays, with AIC$=7.9$, a power law with an exponential cutoff for the particle distribution is preferred over the broken power law with exponential cutoff. The free parameters in the fit were the normalization of the particle distribution, its spectral index and the particle cutoff energy. The resulting index and cutoff energy are $2.11^{+0.08}_{-0.12}$ and $84^{+112}_{-54}$ TeV, respectively. The particle cutoff energy cannot be well constrained using \emph{Fermi}-LAT data alone since the gamma-ray SED itself is best described by a simple power law function. The total energy required in the cosmic rays is $8.5\times 10^{49} \left(\frac{1 \,\mbox{cm}^{-3}}{n} \right) \left(\frac{d}{2\,\mbox{kpc}} \right)^2$ erg, where $n$ is the particle number density of the target material for hadronic interactions and $d$ the distance to the source. The value has been normalized to a sample distance to the source of 2 kpc but it could be easily adjusted for other values. Fig. \ref{fig4:sed} (a) shows the resulting fit to the data in this scenario.

For the IC-dominated scenario we show in Fig. \ref{fig4:sed} (b) the fit to the radio and gamma-ray fluxes using a broken power law with an exponential cutoff (AIC$=14.3$) for the underlying electron distribution, although the fit quality using a power law with an exponential cutoff (AIC$=14.7$) is very similar. The parameters that were free in the fit are the normalization of the particle distribution, the break energy, the spectral indices before and above the break energy, the cutoff energy and the average magnetic field. The spectral index below the break is $1.75^{+0.19}_{-0.14}$, which is consistent with the slope observed in the radio SED, the obtained break energy and index above the break are $23.8^{+12}_{-6.4}$ GeV and $2.88^{+0.22}_{-0.17}$, respectively, while the best-fit cutoff energy and magnetic field are $31^{+12}_{-22}$ TeV and $3.4^{+1.5}_{-0.8}$ $\mu$G, respectively. The total energy in the leptons amounts to $3.6\times 10^{48} \left(\frac{d}{2\,\mbox{kpc}} \right)^2$ erg.

In the case of the bremsstrahlung-dominated scenario the resulting fit has the lowest quality of the scenarios explored. The plot shown in Fig. \ref{fig4:sed} (c) results from the fit with a power law with an exponential cutoff for the particle distribution (having AIC$=92$). The fit has convergence problems if the normalization of the distribution, the magnetic field and the density of the target material are all left free simultaneously, and we chose to fix the ambient density to $n=70$ cm$^{-3}$, which is a plausible value according to Table \ref{tab2:density} for the near kinematic distances to the source. The required spectral index and cutoff energy in the particle distribution are $2.03^{+0.04}_{-0.03}$ and $3.2^{+0.9}_{-0.5}$ TeV, respectively, while the magnetic field obtained is $60^{+26}_{-10}$ $\mu$G. For these parameters and the value used for $n$, the required total energy in the leptons is $\sim 10^{47} \left(\frac{d}{2\,\mbox{kpc}} \right)^2$ erg.

Based only on the quality of the spectral fits to the radio and gamma-ray SED we conclude that if the GeV emission is produced mainly by a single particle population and one physical mechanism is dominant, then the hadronic and IC scenarios are more likely than the bremsstrahlung-dominated scenario.

\subsection{Origin of the gamma-ray emission}
In this section we discuss the possible physical systems that could be responsible for producing extended emission at GeV energies. We note that the gamma-ray emission was reported as steady in time in the 4FGL-DR2 catalog, where the variability indices for 4FGL J1925.4+1616 and 4FGL J1924.3+1628 were estimated as 6.4 and 4.7, respectively \citep{2020arXiv200511208B}. A source is considered variable when this index is above 21.7, corresponding to a 99\%-confidence level in a $\chi^2$ distribution with 9 degrees of freedom \citep[see, e.g.,][]{2020ApJS..247...33A}. The following scenarios are consistent with the steady nature of the emission.

\subsubsection{A star-forming region}
Several young star clusters are known to accelerate cosmic rays most likely at the shocks of the collective winds from massive stars, resulting in the emission of high-energy photons from the matter that surrounds the clusters \citep[see][and references therein]{2019NatAs...3..561A}. The gamma-ray emission associated to these objects is typically extended, with hard GeV spectral indices ($\Gamma \sim 2.1-2.3$) and likely hadronic in origin. In fact, star-forming regions with massive stars may very well constitute at least one class of long sought sources of PeV cosmic rays in the Galaxy \citep{2019NatAs...3..561A}.

\begin{figure*}
   \centering
     \subfigure[]{\includegraphics[width=0.45\textwidth]{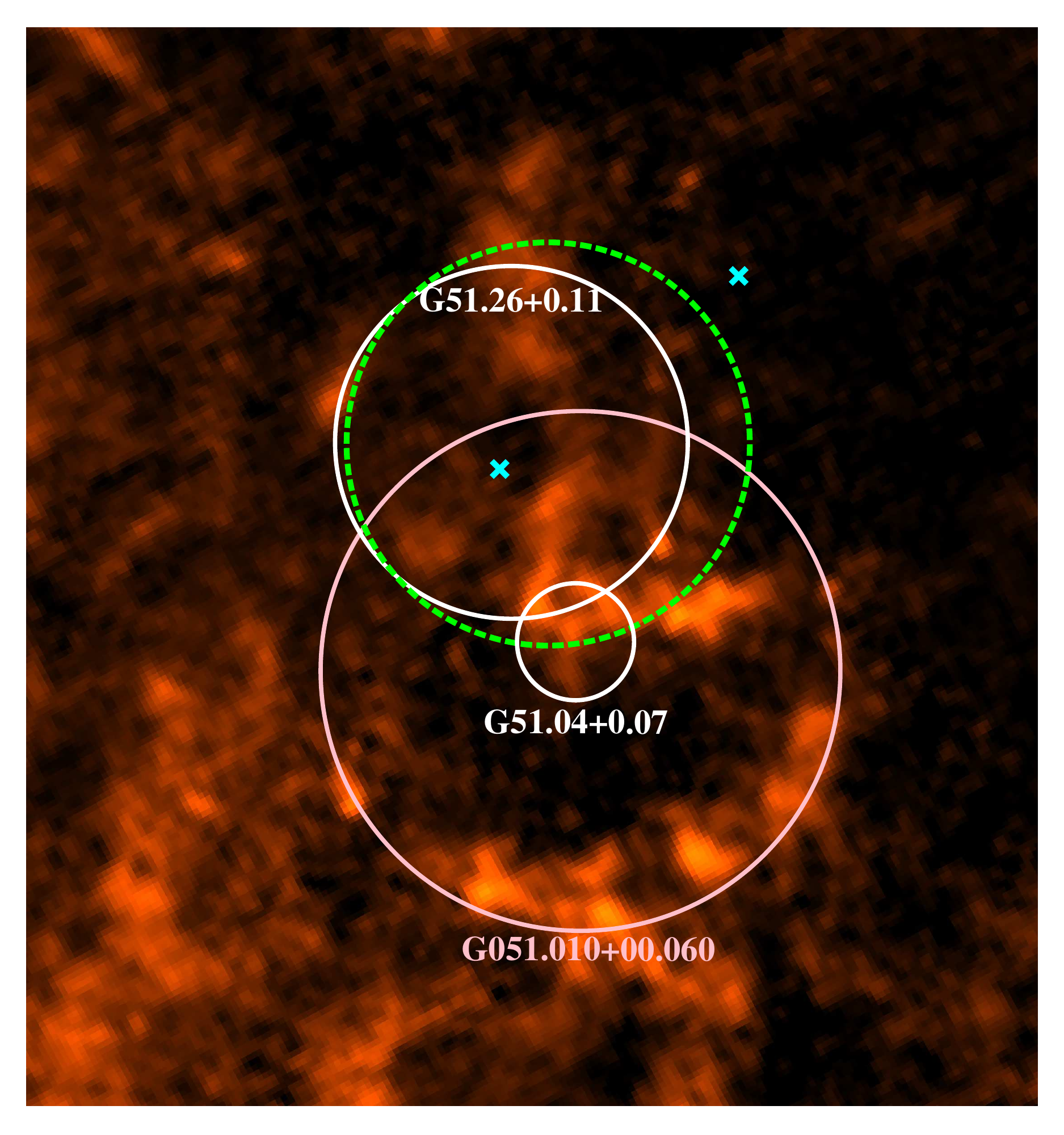}}
     \subfigure[]{\includegraphics[width=0.45\textwidth]{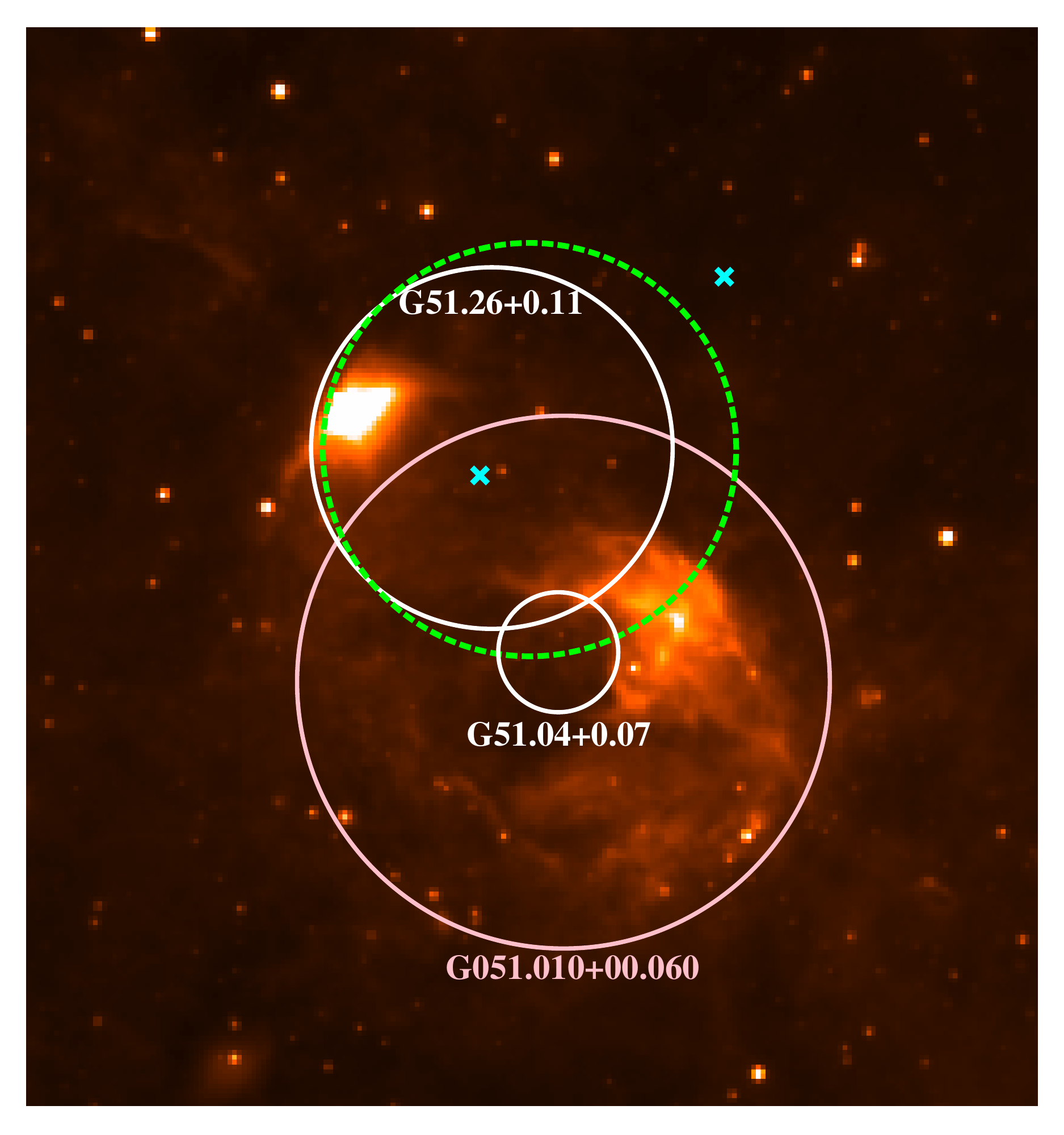}}
    
   \caption{(a) Integrated $^{13}$CO (J = $1\to 0$) intensity in the velocity range 43.4 to 52.2 km s$^{-1}$ from Fig. \ref{fig2:CO} showing the molecular gas associated to the HII region G051.010+00.060. (b) An image of the 22 $\mu$m infrared emission taken from the Wide-field Infrared Survey Explorer \citep{2010AJ....140.1868W} in the region of G51.26+0.11 and G051.010+00.060. Infrared emission around this wavelength in HII regions is characteristic of heated dust. In both panels, the SNRs, the HII region and the LAT sources are shown as in Fig. \ref{fig1:tsmap}.}\label{fig5:HIICO}
\end{figure*}

The most prominent known HII region that partially overlaps the 68\%-containment gamma-ray region seen in Fig. \ref{fig1:tsmap} is G051.010+00.060, located at a distance of $7.2\pm1.2$ kpc \citep{2014ApJS..212....1A_wise_catalog}. This distance is compatible with the gas emission seen at CO line velocities in the interval from 43.4 to 52.2 km s$^{-1}$. Fig. \ref{fig5:HIICO} shows the integrated $^{13}$CO (J = $1\to 0$) emission in the velocity range from 43.4 to 52.2 km s$^{-1}$ and the mid-infrared emission in the region \citep[from][]{2010AJ....140.1868W}, in comparison to the GeV emission from Fig. \ref{fig1:tsmap}. It can be seen that the molecular gas in the southern part of the HII region is displaced from the GeV emission, as is the molecular gas seen to the west of the location of the compact SNR G51.04+0.07. Infrared emission is particularly prominent also to the west of G51.04+0.07. The emission around $\sim20\,\mu$m from HII regions is associated to heated dust \citep{2011ApJS..194...32A}, thus tracing another component related to the stellar activity. In Fig. \ref{fig5:HIICO} it is seen that the heated gas is also located far away from the peak of the gamma rays, while the GeV emission detected from other star-forming regions is correlated with the gas component. Based on the displacement of the gamma-ray emission from the most prominent cold gas structures as well as the heated gas in the region, we conclude that stars are likely not the source of the GeV photons. We also note that the star-forming regions which are known to produce gamma rays all contain very hot and very massive stars such as O class stars or Wolf-Rayet stars \citep{2020A&A...639A..80S}. The Galactic O-Star Spectroscopic Survey \citep{2016ApJS..224....4M} contains no such objects within the region of the sky shown in Fig. \ref{fig5:HIICO}.

Infrared emission is also apparent to the east of the SNR G51.26+0.11. This emission might be related to the candidate star-forming region G051.457-00.286 \citep[as labeled in the WISE catalog,][]{2014ApJS..212....1A_wise_catalog}. The radio emission from this part of the region is indeed thermal \citep{2018ApJ...866...61D}. Again there are no known massive stars at the location of the eastern infrared emission, making this candidate star-forming region unlikely to accelerate particles that could produce gamma rays. Furthermore, Fig. \ref{fig2:CO} shows a molecular cloud coincident with this infrared enhancement in the velocity interval $53-58.8$ km s$^{-1}$. This gas would serve as target material for hadronic interactions producing gamma rays, but the location of the gamma-ray emission is displaced from the dense gas, also making the association unlikely.

\subsubsection{An SNR}
SNRs are known to produce gamma rays. It becomes obvious to propose a possible relation between the GeV source and the SNR G51.26+0.11. As can be seen in Fig. \ref{fig1:tsmap} the gamma-ray emission has a peak near the center of the SNR and its extension is slightly larger than the known extension of G51.26+0.11. We could speculate that the SNR shell is actually larger than proposed recently, given that there is radio continuum emission towards the northwest as revealed by THOR+VGPS data. It is unknown at this point if this radio emission is mostly non thermal, and thus produced by an SNR (either G51.26+0.11 or a different object), or not \citep[see Fig. \ref{fig1:tsmap} and also the maps shown by][]{2017A&A...605A..58A_THOR,2018A&A...619A.124W}. The much more compact SNR G51.04+0.07 described by \cite{2018A&A...616A..98S} is seen outside the GeV emission region and therefore is not considered here as a source of the gamma rays.

Regarding the CO emission seen in Fig. \ref{fig2:CO}, a small cloud of molecular gas is seen inside the projected shell of the SNR in the $15.2-18.6$ km s$^{-1}$ velocity range. According to our estimates in Table \ref{tab2:density} this cloud could have sufficient density for hadronic interactions, however it occupies only a small portion of the projected area of the GeV source (which is much more extended than the cloud) and the peak of the gamma ray emission is displaced from the location of this cloud. This makes the association between the gas in the velocity interval $15.2-18.6$ km s$^{-1}$ and the GeV source unlikely.

For the other cases where molecular gas is detected, the velocity interval $53-58.8$ km s$^{-1}$ contains the highest integrated emission as well as gas with considerable density at the location of the peak of the gamma rays, a correlation which would be expected in a hadronic scenario for the GeV photons. This gas and the gas seen in the velocity interval $43.4-52.2$ km s$^{-1}$ seem to fill a region comparable to that of the GeV source, and the latter also has enough density for hadronic interactions. It is then possible that the SNR G51.26+0.11 is a source of high-energy particles which illuminate a nearby or interacting cloud of gas, and the system would be located $\sim3.7 - 6.6$ kpc from the observer (see Table \ref{tab2:density}).

For a distance of 3.7 (6.6) kpc to the source, according to our SED fit in the hadronic scenario and our estimated density values in Table \ref{tab2:density}, the required total cosmic ray energy would be $3.5\times 10^{48}$ erg ($2\times 10^{49}$ erg), which is much lower than the typical available kinetic energy in an SNR ($\sim 10^{51}$ erg). For a distance of 5 kpc and a hydrogen number density of 80 cm$^{-3}$ (corresponding to the gas in the velocity interval $53-58.8$ km s$^{-1}$), the total energy in the cosmic rays is $6.6\times 10^{48}$ erg, which is also reasonable.

It is important to point out that our conclusion regarding the feasibility of the hadronic model where cosmic rays from the SNR interact with gas seen at either of the plausible velocity intervals ($43.4-52.2$ and $53-58.8$ km s$^{-1}$) is not really affected by any uncertainties in the model used to estimate the cloud distances (at least from the point of view of the energy required). Even for extreme distances to the clouds of 15 kpc, after properly calculating the corresponding gas densities, the required total energy in the cosmic rays would be $\sim$20\% of the typical kinetic energy available in an SNR.

The extension of the GeV source found here (taken as the full width at half maximum for the Gaussian found, $\sim 0.34^\circ$) is consistent with the diameter of the SNR shell \citep[$\sim 0.38^\circ$ according to][]{2018ApJ...866...61D}. If $0.38^\circ$ is taken as the real extent of the SNR shell, this would translate to a physical diameter of $6.6 \left( \frac{d}{1\,\mbox{kpc}} \right)$ pc, where $d$ is the source distance. The diameters of known SNRs are typically of the order of parsecs to several tens of parsecs, and therefore the observed source extension could be consistent with that of an SNR for the range of distances estimated to the gas of $3.7-6.6$ kpc.

The luminosity observed in the gamma rays is also common in SNRs for a large range of possible distances to the source. From the photon spectrum determined in Section \ref{sec:lat}, we estimated a luminosity of the GeV emission in the $1-500$ GeV energy interval, for a distance $d$ to the source, of $1.2\times 10^{34} \left( \frac{d}{2\, \mbox{kpc}} \right)^2$ erg s$^{-1}$. This value is typical of GeV-emitting SNRs \citep{2016ApJS..224....8A}.

SNRs also show a variety of GeV spectral indices and shapes, thought to be caused by the differences in environments and evolutionary stages. With a spectral index of $\sim 2.2$, the high-energy spectrum of the source shows a similar shape to that of Cas A's in the energy range 1--500 GeV \citep[with a spectral index of $2.17\pm 0.02_{\mbox{\tiny stat}}$,][]{2020ApJ...894...51A,2013ApJ...779..117Y} and that of Tycho's SNR in the GeV range \citep[with a spectral index of $2.14\pm 0.09_{\mbox{\tiny stat}}$,][]{2017ApJ...836...23A}. Both of these are young SNRs with bright X-ray and radio emission. We encourage carrying out observations at X-ray energies in the region of G51.26+0.11 to help constrain the parameters of the SNR.


The hadronic interpretation of the high-energy SED is consistent with an SNR scenario for the gamma rays and the ISM properties presented in this work. The hadronic scenario also provides the best fit to the GeV data, as shown before. The leptonic scenarios presented here pose no problem from the point of view of the energetics, as a typical SNR kinetic energy value is more than enough to channel the required energy to the particles. The order of magnitude of the total energy in the particles needed in any scenario is consistent with measured values for SNRs \citep[see, e.g.,][]{2010ApJ...720...20A,2016ApJ...826...31F,2019PASJ...71...77X}. In terms of the fit quality, the bremsstrahlung scenario shows some tension with the measured radio fluxes in the simple, one-zone, approximation used here. A more detailed modeling considering multiple emission zones is left for the future.

\subsubsection{A pulsar wind nebula}
Pulsars can also produce extended gamma-ray radiation through the interaction of relativistic electrons and positrons, accelerated by the pulsar wind nebula (PWN), with ambient photons. A scenario where an SNR produces the observed radio emission and contains within it a PWN that shines in the gamma-ray range seems plausible. This would also explain the Gaussian-like morphology of the GeV emission having a peak towards the center of G51.26+0.11.

A search for pulsars from the Australia Telescope National Facility Pulsar Catalogue \citep{2005AJ....129.1993M}\footnote{https://www.atnf.csiro.au/research/pulsar/psrcat/} within $0.4^\circ$ of the peak of the gamma-ray emission yields three pulsars, PSR J1924+1628, PSR J1924+1631 and PSR J1926+1613. The first two have catalogued distances of 10.7 and 10.2 kpc, respectively, and are seen outside the 68\%-containment region of the gamma-ray source shown in Fig. \ref{fig1:tsmap}. PSR J1924+1631 has a relatively low spin-down power ($5.7\times10^{32}$ erg s$^{-1}$), three orders of magnitude lower than the observed gamma-ray luminosity for the same distance, and thus we consider it an unlikely source of the gamma rays. The energy required in the particles according to our IC-dominated scenario, which is typical in a PWN, is $\sim9\times 10^{49}$ erg for a distance of 10 kpc, and cannot be provided by a pulsar with such low spin-down power. The spin-down power of PSR J1924+1628 is not known, however, its distance of 10.7 kpc would make the GeV PWN unusually large \citep[although comparable to the GeV size of the PWN HESS J1825-137,][]{2020A&A...640A..76P}.

On the other hand, the pulsar PSR J1926+1613 is seen $\sim0.2^\circ$ away from the peak of the gamma-ray emission (near the edge of the LAT 68\%-containment circle). This pulsar is located at a distance of 1.5 kpc, but its parameters such as spin-down power and age are unknown. We cannot confirm or rule out this source as the origin of the high-energy emission. X-ray observations of G51.26+0.11 would be helpful to identify any possible PWN in the region, either associated to PSR J1926+1613 or to any other currently undiscovered pulsar.


%

\section{Conclusions}
We have found an extended gamma-ray source with a hard spectrum (photon index $\sim2.18$) in the GeV range using data from the \emph{Fermi} satellite at the location of the SNR G51.26+0.11. Based on the radio and gamma-ray SED, one-zone scenarios involving hadronic emission from cosmic rays and leptonic emission from IC scattering of background photons by electrons are both possible, with the hadronic scenario providing the best fit to the GeV fluxes. A bremsstrahlung-dominated (one-zone) scenario for the gamma-rays provides the worst fit to the broadband data.

We rule out the star-forming region G051.010+00.060 as the source of the high-energy photons based on the lack of correspondence between the gas in this region and the GeV emission, as well as the absence of massive stars. This is also true for the candidate star-forming region G051.457-00.286.

The SNR G51.26+0.11 is a likely source of particles producing the gamma rays. The GeV emission is seen at the location of this SNR, although it is slightly more extended than the proposed radio shell. The hadronic scenario for the gamma rays is compatible with molecular gas clouds seen at the location of the SNR/GeV source in two velocity intervals obtained from CO observations, $43.4-52.2$ and $53-58.8$ km s$^{-1}$. This would place the SNR at a distance in the range of $3.7-6.6$ kpc, but the conclusion regarding the feasibility of the hadronic model is not affected by changing the distance to the source to any other reasonable value. From the point of view of the energetics, the required total energy in the relativistic particles can be supplied by a typical supernova event, in both hadronic and leptonic scenarios, for any reasonable distance to the source.

The gamma-ray SED can also be well explained by IC emission from leptons, which is thought to dominate in the gamma-ray emission of some SNRs as well as in PWN. Given the lack of parameters it is unclear whether the pulsar PSR J1926+1613 could be responsible for the high-energy emission. A search for new pulsars or an X-ray PWN (perhaps associated to G51.26+0.11) would be very valuable to confirm the origin of the radiation detected by \emph{Fermi}-LAT.

\begin{acknowledgements}
We thank the anonymous referee for useful comments that improved this work. Thanks to L. Anderson, X. Sun and J. Stil for their comments regarding the radio data. This work was possible due to funding by Universidad de Costa Rica and its Escuela de F\'isica under grant number B8267. This research is based on observations made with NASA's Fermi Gamma-Ray Space Telescope, developed in collaboration with the U.S. Department of Energy, along with important contributions from academic institutions and partners in France, Germany, Italy, Japan, Sweden and the U.S. It also makes use of molecular line data from the Boston University-FCRAO Galactic Ring Survey (GRS). The GRS is a joint project of Boston University and Five College Radio Astronomy Observatory, funded by the National Science Foundation under grants AST-9800334, AST-0098562, AST-0100793, AST-0228993 and AST-0507657.

\end{acknowledgements}

\bibliographystyle{aasjournal}
\bibliography{references}{}

\begin{thebibliography}{}
\expandafter\ifx\csname natexlab\endcsname\relax\def\natexlab#1{#1}\fi
\providecommand{\url}[1]{\href{#1}{#1}}
\providecommand{\dodoi}[1]{doi:~\href{http://doi.org/#1}{\nolinkurl{#1}}}
\providecommand{\doeprint}[1]{\href{http://ascl.net/#1}{\nolinkurl{http://ascl.net/#1}}}
\providecommand{\doarXiv}[1]{\href{https://arxiv.org/abs/#1}{\nolinkurl{https://arxiv.org/abs/#1}}}

\bibitem[{{Abdollahi} {et~al.}(2020){Abdollahi}, {Acero}, {Ackermann},
  {Ajello}, {Atwood}, {Axelsson}, {Baldini}, {Ballet}, {Barbiellini},
  {Bastieri}, {Becerra Gonzalez}, {Bellazzini}, {Berretta}, {Bissaldi},
  {Blandford}, {Bloom}, {Bonino}, {Bottacini}, {Brandt}, {Bregeon}, {Bruel},
  {Buehler}, {Burnett}, {Buson}, {Cameron}, {Caputo}, {Caraveo}, {Casandjian},
  {Castro}, {Cavazzuti}, {Charles}, {Chaty}, {Chen}, {Cheung}, {Chiaro},
  {Ciprini}, {Cohen-Tanugi}, {Cominsky}, {Coronado-Bl{\'a}zquez}, {Costantin},
  {Cuoco}, {Cutini}, {D'Ammando}, {DeKlotz}, {de la Torre Luque}, {de Palma},
  {Desai}, {Digel}, {Di Lalla}, {Di Mauro}, {Di Venere}, {Dom{\'\i}nguez},
  {Dumora}, {Fana Dirirsa}, {Fegan}, {Ferrara}, {Franckowiak}, {Fukazawa},
  {Funk}, {Fusco}, {Gargano}, {Gasparrini}, {Giglietto}, {Giommi}, {Giordano},
  {Giroletti}, {Glanzman}, {Green}, {Grenier}, {Griffin}, {Grondin}, {Grove},
  {Guiriec}, {Harding}, {Hayashi}, {Hays}, {Hewitt}, {Horan},
  {J{\'o}hannesson}, {Johnson}, {Kamae}, {Kerr}, {Kocevski}, {Kovac'evic'},
  {Kuss}, {Landriu}, {Larsson}, {Latronico}, {Lemoine-Goumard}, {Li},
  {Liodakis}, {Longo}, {Loparco}, {Lott}, {Lovellette}, {Lubrano}, {Madejski},
  {Maldera}, {Malyshev}, {Manfreda}, {Marchesini}, {Marcotulli},
  {Mart{\'\i}-Devesa}, {Martin}, {Massaro}, {Mazziotta}, {McEnery}, {Mereu},
  {Meyer}, {Michelson}, {Mirabal}, {Mizuno}, {Monzani}, {Morselli},
  {Moskalenko}, {Negro}, {Nuss}, {Ojha}, {Omodei}, {Orienti}, {Orlando},
  {Ormes}, {Palatiello}, {Paliya}, {Paneque}, {Pei}, {Pe{\~n}a-Herazo},
  {Perkins}, {Persic}, {Pesce-Rollins}, {Petrosian}, {Petrov}, {Piron}, {Poon},
  {Porter}, {Principe}, {Rain{\`o}}, {Rando}, {Razzano}, {Razzaque}, {Reimer},
  {Reimer}, {Remy}, {Reposeur}, {Romani}, {Saz Parkinson}, {Schinzel},
  {Serini}, {Sgr{\`o}}, {Siskind}, {Smith}, {Spandre}, {Spinelli}, {Strong},
  {Suson}, {Tajima}, {Takahashi}, {Tak}, {Thayer}, {Thompson}, {Tibaldo},
  {Torres}, {Torresi}, {Valverde}, {Van Klaveren}, {van Zyl}, {Wood},
  {Yassine}, \& {Zaharijas}}]{2020ApJS..247...33A}
{Abdollahi}, S., {Acero}, F., {Ackermann}, M., {et~al.} 2020, \apjs, 247, 33,
  \dodoi{10.3847/1538-4365/ab6bcb}

\bibitem[{{Abeysekara} {et~al.}(2020){Abeysekara}, {Archer}, {Benbow}, {Bird},
  {Brose}, {Buchovecky}, {Buckley}, {Chromey}, {Cui}, {Daniel}, {Das},
  {Dwarkadas}, {Falcone}, {Feng}, {Finley}, {Fortson}, {Gent}, {Gillanders},
  {Giuri}, {Gueta}, {Hanna}, {Hassan}, {Hervet}, {Holder}, {Hughes},
  {Humensky}, {Kaaret}, {Kar}, {Kelley-Hoskins}, {Kertzman}, {Kieda}, {Krause},
  {Krennrich}, {Kumar}, {Lang}, {Maier}, {Moriarty}, {Mukherjee},
  {Nievas-Rosillo}, {O'Brien}, {Ong}, {Park}, {Petrashyk}, {Pfrang}, {Pohl},
  {Pueschel}, {Quinn}, {Ragan}, {Reynolds}, {Richards}, {Roache}, {Sadeh},
  {Santander}, {Sembroski}, {Shahinyan}, {Sushch}, {Weinstein}, {Wilcox},
  {Wilhelm}, {Williams}, {Williamson}, {Zitzer}, \&
  {Ghiotto}}]{2020ApJ...894...51A}
{Abeysekara}, A.~U., {Archer}, A., {Benbow}, W., {et~al.} 2020, \apj, 894, 51,
  \dodoi{10.3847/1538-4357/ab8310}

\bibitem[{{Acero} {et~al.}(2016){Acero}, {Ackermann}, {Ajello}, {Baldini},
  {Ballet}, {Barbiellini}, {Bastieri}, {Bellazzini}, {Bissaldi}, {Blandford},
  {Bloom}, {Bonino}, {Bottacini}, {Brandt}, {Bregeon}, {Bruel}, {Buehler},
  {Buson}, {Caliandro}, {Cameron}, {Caputo}, {Caragiulo}, {Caraveo},
  {Casandjian}, {Cavazzuti}, {Cecchi}, {Chekhtman}, {Chiang}, {Chiaro},
  {Ciprini}, {Claus}, {Cohen}, {Cohen-Tanugi}, {Cominsky}, {Condon}, {Conrad},
  {Cutini}, {D'Ammando}, {de Angelis}, {de Palma}, {Desiante}, {Digel}, {Di
  Venere}, {Drell}, {Drlica-Wagner}, {Favuzzi}, {Ferrara}, {Franckowiak},
  {Fukazawa}, {Funk}, {Fusco}, {Gargano}, {Gasparrini}, {Giglietto}, {Giommi},
  {Giordano}, {Giroletti}, {Glanzman}, {Godfrey}, {Gomez-Vargas}, {Grenier},
  {Grondin}, {Guillemot}, {Guiriec}, {Gustafsson}, {Hadasch}, {Harding},
  {Hayashida}, {Hays}, {Hewitt}, {Hill}, {Horan}, {Hou}, {Iafrate}, {Jogler},
  {J{\'o}hannesson}, {Johnson}, {Kamae}, {Katagiri}, {Kataoka}, {Katsuta},
  {Kerr}, {Kn{\"o}dlseder}, {Kocevski}, {Kuss}, {Laffon}, {Lande}, {Larsson},
  {Latronico}, {Lemoine-Goumard}, {Li}, {Li}, {Longo}, {Loparco}, {Lovellette},
  {Lubrano}, {Magill}, {Maldera}, {Marelli}, {Mayer}, {Mazziotta}, {Michelson},
  {Mitthumsiri}, {Mizuno}, {Moiseev}, {Monzani}, {Moretti}, {Morselli},
  {Moskalenko}, {Murgia}, {Nemmen}, {Nuss}, {Ohsugi}, {Omodei}, {Orienti},
  {Orlando}, {Ormes}, {Paneque}, {Perkins}, {Pesce-Rollins}, {Petrosian},
  {Piron}, {Pivato}, {Porter}, {Rain{\`o}}, {Rando}, {Razzano}, {Razzaque},
  {Reimer}, {Reimer}, {Renaud}, {Reposeur}, {Rousseau}, {Saz Parkinson},
  {Schmid}, {Schulz}, {Sgr{\`o}}, {Siskind}, {Spada}, {Spandre}, {Spinelli},
  {Strong}, {Suson}, {Tajima}, {Takahashi}, {Tanaka}, {Thayer}, {Thompson},
  {Tibaldo}, {Tibolla}, {Torres}, {Tosti}, {Troja}, {Uchiyama}, {Vianello},
  {Wells}, {Wood}, {Wood}, {Yassine}, {den Hartog}, \&
  {Zimmer}}]{2016ApJS..224....8A}
{Acero}, F., {Ackermann}, M., {Ajello}, M., {et~al.} 2016, \apjs, 224, 8,
  \dodoi{10.3847/0067-0049/224/1/8}

\bibitem[{{Ackermann} {et~al.}(2012){Ackermann}, {Ajello}, {Albert},
  {Allafort}, {Atwood}, {Axelsson}, {Baldini}, {Ballet}, {Barbiellini},
  {Bastieri}, {Bechtol}, {Bellazzini}, {Bissaldi}, {Blandford}, {Bloom},
  {Bogart}, {Bonamente}, {Borgland}, {Bottacini}, {Bouvier}, {Brandt},
  {Bregeon}, {Brigida}, {Bruel}, {Buehler}, {Burnett}, {Buson}, {Caliandro},
  {Cameron}, {Caraveo}, {Casandjian}, {Cavazzuti}, {Cecchi}, {{\c C}elik},
  {Charles}, {Chaves}, {Chekhtman}, {Cheung}, {Chiang}, {Ciprini}, {Claus},
  {Cohen-Tanugi}, {Conrad}, {Corbet}, {Cutini}, {D'Ammando}, {Davis}, {de
  Angelis}, {DeKlotz}, {de Palma}, {Dermer}, {Digel}, {Silva}, {Drell},
  {Drlica-Wagner}, {Dubois}, {Favuzzi}, {Fegan}, {Ferrara}, {Focke}, {Fortin},
  {Fukazawa}, {Funk}, {Fusco}, {Gargano}, {Gasparrini}, {Gehrels}, {Giebels},
  {Giglietto}, {Giordano}, {Giroletti}, {Glanzman}, {Godfrey}, {Grenier},
  {Grove}, {Guiriec}, {Hadasch}, {Hayashida}, {Hays}, {Horan}, {Hou}, {Hughes},
  {Jackson}, {Jogler}, {J{\'o}hannesson}, {Johnson}, {Johnson}, {Johnson},
  {Kamae}, {Katagiri}, {Kataoka}, {Kerr}, {Kn{\"o}dlseder}, {Kuss}, {Lande},
  {Larsson}, {Latronico}, {Lavalley}, {Lemoine-Goumard}, {Longo}, {Loparco},
  {Lott}, {Lovellette}, {Lubrano}, {Mazziotta}, {McConville}, {McEnery},
  {Mehault}, {Michelson}, {Mitthumsiri}, {Mizuno}, {Moiseev}, {Monte},
  {Monzani}, {Morselli}, {Moskalenko}, {Murgia}, {Naumann-Godo}, {Nemmen},
  {Nishino}, {Norris}, {Nuss}, {Ohno}, {Ohsugi}, {Okumura}, {Omodei},
  {Orienti}, {Orlando}, {Ormes}, {Paneque}, {Panetta}, {Perkins},
  {Pesce-Rollins}, {Pierbattista}, {Piron}, {Pivato}, {Porter}, {Racusin},
  {Rain{\`o}}, {Rando}, {Razzano}, {Razzaque}, {Reimer}, {Reimer}, {Reposeur},
  {Reyes}, {Ritz}, {Rochester}, {Romoli}, {Roth}, {Sadrozinski}, {Sanchez},
  {Saz Parkinson}, {Sbarra}, {Scargle}, {Sgr{\`o}}, {Siegal-Gaskins},
  {Siskind}, {Spandre}, {Spinelli}, {Stephens}, {Suson}, {Tajima}, {Takahashi},
  {Tanaka}, {Thayer}, {Thayer}, {Thompson}, {Tibaldo}, {Tinivella}, {Tosti},
  {Troja}, {Usher}, {Vandenbroucke}, {Van Klaveren}, {Vasileiou}, {Vianello},
  {Vitale}, {Waite}, {Wallace}, {Winer}, {Wood}, {Wood}, {Wood}, {Yang}, \&
  {Zimmer}}]{2012ApJS..203....4A}
{Ackermann}, M., {Ajello}, M., {Albert}, A., {et~al.} 2012, \apjs, 203, 4,
  \dodoi{10.1088/0067-0049/203/1/4}

\bibitem[{{Aharonian} {et~al.}(2019){Aharonian}, {Yang}, \& {de O{\~n}a
  Wilhelmi}}]{2019NatAs...3..561A}
{Aharonian}, F., {Yang}, R., \& {de O{\~n}a Wilhelmi}, E. 2019, Nature
  Astronomy, 3, 561, \dodoi{10.1038/s41550-019-0724-0}

\bibitem[{{Akaike}(1974)}]{1974ITAC...19..716A}
{Akaike}, H. 1974, IEEE Transactions on Automatic Control, 19, 716

\bibitem[{{Albert} {et~al.}(2020){Albert}, {Alfaro}, {Alvarez}, {Camacho},
  {Arteaga-Vel{\'a}zquez}, {Arunbabu}, {Avila Rojas}, {Ayala Solares},
  {Baghmanyan}, {Belmont-Moreno}, {BenZvi}, {Brisbois}, {Caballero-Mora},
  {Capistr{\'a}n}, {Carrami{\~n}ana}, {Casanova}, {Cotti}, {Couti{\~n}o de
  Le{\'o}n}, {De la Fuente}, {Diaz Hernandez}, {Diaz-Cruz}, {Dingus},
  {DuVernois}, {Durocher}, {D{\'\i}az-V{\'e}lez}, {Ellsworth}, {Engel},
  {Espinoza}, {Fan}, {Fang}, {Alonso}, {Fleischhack}, {Fraija},
  {Galv{\'a}n-G{\'a}mez}, {Garcia}, {Garc{\'\i}a-Gonz{\'a}lez}, {Garfias},
  {Giacinti}, {Gonz{\'a}lez}, {Goodman}, {Harding}, {Hernandez}, {Hinton},
  {Hona}, {Huang}, {Hueyotl-Zahuantitla}, {H{\"u}ntemeyer}, {Iriarte},
  {Jardin-Blicq}, {Joshi}, {Kieda}, {Lara}, {Lee}, {Le{\'o}n Vargas},
  {Linnemann}, {Longinotti}, {Luis-Raya}, {Lundeen}, {L{\'o}pez-Coto},
  {Malone}, {Marandon}, {Martinez}, {Martinez-Castellanos},
  {Mart{\'\i}nez-Castro}, {Matthews}, {Miranda-Romagnoli}, {Morales-Soto},
  {Moreno}, {Mostaf{\'a}}, {Nayerhoda}, {Nellen}, {Newbold}, {Nisa},
  {Noriega-Papaqui}, {Olivera-Nieto}, {Omodei}, {Peisker}, {P{\'e}rez Araujo},
  {P{\'e}rez-P{\'e}rez}, {Ren}, {Rho}, {Rivi{\`e}re}, {Rosa-Gonz{\'a}lez},
  {Ruiz-Velasco}, {Salazar}, {Salesa Greus}, {Sandoval}, {Schneider},
  {Schoorlemmer}, {Serna}, {Sinnis}, {Smith}, {Springer}, {Surajbali},
  {Tollefson}, {Torres}, {Torres-Escobedo}, {Ukwatta}, {Ure{\~n}a-Mena},
  {Weisgarber}, {Werner}, {Willox}, {Zepeda}, {Zhou}, {de Le{\'o}n},
  {{\'A}lvarez}, \& {HAWC Collaboration}}]{2020ApJ...905...76A}
{Albert}, A., {Alfaro}, R., {Alvarez}, C., {et~al.} 2020, \apj, 905, 76,
  \dodoi{10.3847/1538-4357/abc2d8}

\bibitem[{{Ambrogi} {et~al.}(2019){Ambrogi}, {Zanin}, {Casanova}, {De O{\~n}a
  Wilhelmi}, {Peron}, \& {Aharonian}}]{2019A&A...623A..86A}
{Ambrogi}, L., {Zanin}, R., {Casanova}, S., {et~al.} 2019, \aap, 623, A86,
  \dodoi{10.1051/0004-6361/201833985}

\bibitem[{Anderson {et~al.}(2014)Anderson, Bania, Balser, Cunningham, Wenger,
  Johnstone, \& Armentrout}]{Anderson_2014}
Anderson, L.~D., Bania, T.~M., Balser, D.~S., {et~al.} 2014, The Astrophysical
  Journal Supplement Series, 212, 1, \dodoi{10.1088/0067-0049/212/1/1}

\bibitem[{{Anderson} {et~al.}(2014){Anderson}, {Bania}, {Balser}, {Cunningham},
  {Wenger}, {Johnstone}, \& {Armentrout}}]{2014ApJS..212....1A_wise_catalog}
{Anderson}, L.~D., {Bania}, T.~M., {Balser}, D.~S., {et~al.} 2014, \apjs, 212,
  1, \dodoi{10.1088/0067-0049/212/1/1}

\bibitem[{{Anderson} {et~al.}(2011){Anderson}, {Bania}, {Balser}, \&
  {Rood}}]{2011ApJS..194...32A}
{Anderson}, L.~D., {Bania}, T.~M., {Balser}, D.~S., \& {Rood}, R.~T. 2011,
  \apjs, 194, 32, \dodoi{10.1088/0067-0049/194/2/32}

\bibitem[{{Anderson} {et~al.}(2017){Anderson}, {Wang}, {Bihr}, {Rugel},
  {Beuther}, {Bigiel}, {Churchwell}, {Glover}, {Goodman}, {Henning}, {Heyer},
  {Klessen}, {Linz}, {Longmore}, {Menten}, {Ott}, {Roy}, {Soler}, {Stil}, \&
  {Urquhart}}]{2017A&A...605A..58A_THOR}
{Anderson}, L.~D., {Wang}, Y., {Bihr}, S., {et~al.} 2017, \aap, 605, A58,
  \dodoi{10.1051/0004-6361/201731019}

\bibitem[{{Araya} \& {Cui}(2010)}]{2010ApJ...720...20A}
{Araya}, M., \& {Cui}, W. 2010, \apj, 720, 20,
  \dodoi{10.1088/0004-637X/720/1/20}

\bibitem[{{Archambault} {et~al.}(2017){Archambault}, {Archer}, {Benbow},
  {Bird}, {Bourbeau}, {Buchovecky}, {Buckley}, {Bugaev}, {Cerruti}, {Connolly},
  {Cui}, {Dwarkadas}, {Errando}, {Falcone}, {Feng}, {Finley}, {Fleischhack},
  {Fortson}, {Furniss}, {Griffin}, {H{\"u}tten}, {Hanna}, {Holder}, {Johnson},
  {Kaaret}, {Kar}, {Kelley-Hoskins}, {Kertzman}, {Kieda}, {Krause}, {Kumar},
  {Lang}, {Maier}, {McArthur}, {McCann}, {Moriarty}, {Mukherjee}, {Nieto},
  {O'Brien}, {Ong}, {Otte}, {Park}, {Pohl}, {Popkow}, {Pueschel}, {Quinn},
  {Ragan}, {Reynolds}, {Richards}, {Roache}, {Sadeh}, {Santander}, {Sembroski},
  {Shahinyan}, {Slane}, {Staszak}, {Telezhinsky}, {Trepanier}, {Tyler},
  {Wakely}, {Weinstein}, {Weisgarber}, {Wilcox}, {Wilhelm}, {Williams}, \&
  {Zitzer}}]{2017ApJ...836...23A}
{Archambault}, S., {Archer}, A., {Benbow}, W., {et~al.} 2017, \apj, 836, 23,
  \dodoi{10.3847/1538-4357/836/1/23}

\bibitem[{{Ballet} {et~al.}(2020){Ballet}, {Burnett}, {Digel}, \&
  {Lott}}]{2020arXiv200511208B}
{Ballet}, J., {Burnett}, T.~H., {Digel}, S.~W., \& {Lott}, B. 2020, arXiv
  e-prints, arXiv:2005.11208.
\newblock \doarXiv{2005.11208}

\bibitem[{{Beuther} {et~al.}(2016){Beuther}, {Bihr}, {Rugel}, {Johnston},
  {Wang}, {Walter}, {Brunthaler}, {Walsh}, {Ott}, {Stil}, {Henning},
  {Schierhuber}, {Kainulainen}, {Heyer}, {Goldsmith}, {Anderson}, {Longmore},
  {Klessen}, {Glover}, {Urquhart}, {Plume}, {Ragan}, {Schneider},
  {McClure-Griffiths}, {Menten}, {Smith}, {Roy}, {Shanahan}, {Nguyen-Luong}, \&
  {Bigiel}}]{2016A&A...595A..32B}
{Beuther}, H., {Bihr}, S., {Rugel}, M., {et~al.} 2016, \aap, 595, A32,
  \dodoi{10.1051/0004-6361/201629143}

\bibitem[{{Brand} \& {Blitz}(1993)}]{1993A&A...275...67B}
{Brand}, J., \& {Blitz}, L. 1993, \aap, 275, 67

\bibitem[{{Dame} {et~al.}(2001){Dame}, {Hartmann}, \&
  {Thaddeus}}]{2001ApJ...547..792D}
{Dame}, T.~M., {Hartmann}, D., \& {Thaddeus}, P. 2001, \apj, 547, 792,
  \dodoi{10.1086/318388}

\bibitem[{{Dokara} {et~al.}(2018){Dokara}, {Roy}, {Beuther}, {Anderson},
  {Rugel}, {Stil}, {Wang}, {Soler}, \& {Shanahan}}]{2018ApJ...866...61D}
{Dokara}, R., {Roy}, N., {Beuther}, H., {et~al.} 2018, \apj, 866, 61,
  \dodoi{10.3847/1538-4357/aadc0c}

\bibitem[{{Dokara} {et~al.}(2021){Dokara}, {Brunthaler}, {Menten}, {Dzib},
  {Reich}, {Cotton}, {Anderson}, {Chen}, {Gong}, {Medina}, {Ortiz-Le{\'o}n},
  {Rugel}, {Urquhart}, {Wyrowski}, {Yang}, {Beuther}, {Billington}, {Csengeri},
  {Carrasco-Gonz{\'a}lez}, \& {Roy}}]{2021arXiv210306267D}
{Dokara}, R., {Brunthaler}, A., {Menten}, K.~M., {et~al.} 2021, arXiv e-prints,
  arXiv:2103.06267.
\newblock \doarXiv{2103.06267}

\bibitem[{{Driessen} {et~al.}(2018){Driessen}, {Dom{\v{c}}ek}, {Vink},
  {Hessels}, {Arias}, \& {Gelfand}}]{2018ApJ...860..133D}
{Driessen}, L.~N., {Dom{\v{c}}ek}, V., {Vink}, J., {et~al.} 2018, \apj, 860,
  133, \dodoi{10.3847/1538-4357/aac32e}

\bibitem[{{Fraija} \& {Araya}(2016)}]{2016ApJ...826...31F}
{Fraija}, N., \& {Araya}, M. 2016, \apj, 826, 31,
  \dodoi{10.3847/0004-637X/826/1/31}

\bibitem[{{Green}(2019)}]{2019JApA...40...36G}
{Green}, D.~A. 2019, Journal of Astrophysics and Astronomy, 40, 36,
  \dodoi{10.1007/s12036-019-9601-6}

\bibitem[{{H.~E.~S.~S. Collaboration} {et~al.}(2015){H.~E.~S.~S.
  Collaboration}, {Abramowski}, {Aharonian}, {Ait Benkhali}, {Akhperjanian},
  {Ang{\"u}ner}, {Backes}, {Balenderan}, {Balzer}, {Barnacka}, {Becherini},
  {Becker Tjus}, {Berge}, {Bernhard}, {Bernl{\"o}hr}, {Birsin}, {Biteau},
  {B{\"o}ttcher}, {Boisson}, {Bolmont}, {Bordas}, {Bregeon}, {Brun}, {Brun},
  {Bryan}, {Bulik}, {Carrigan}, {Casanova}, {Chadwick}, {Chakraborty},
  {Chalme-Calvet}, {Chaves}, {Chr{\'e}tien}, {Colafrancesco}, {Cologna},
  {Conrad}, {Couturier}, {Cui}, {Davids}, {Degrange}, {Deil}, {deWilt},
  {Djannati-Ata{\"\i}}, {Domainko}, {Donath}, {O'C. Drury}, {Dubus}, {Dutson},
  {Dyks}, {Dyrda}, {Edwards}, {Egberts}, {Eger}, {Espigat}, {Farnier}, {Fegan},
  {Feinstein}, {Fernandes}, {Fernandez}, {Fiasson}, {Fontaine}, {F{\"o}rster},
  {F{\"u}{\ss}ling}, {Gabici}, {Gajdus}, {Gallant}, {Garrigoux}, {Giavitto},
  {Giebels}, {Glicenstein}, {Gottschall}, {Grondin}, {Grudzi{\'n}ska},
  {Hadasch}, {H{\"a}ffner}, {Hahn}, {Harris}, {Heinzelmann}, {Henri},
  {Hermann}, {Hervet}, {Hillert}, {Hinton}, {Hofmann}, {Hofverberg}, {Holler},
  {Horns}, {Ivascenko}, {Jacholkowska}, {Jahn}, {Jamrozy}, {Janiak},
  {Jankowsky}, {Jung-Richardt}, {Kastendieck}, {Katarzy{\'n}ski}, {Katz},
  {Kaufmann}, {Kh{\'e}lifi}, {Kieffer}, {Klepser}, {Klochkov}, {Klu{\'z}niak},
  {Kolitzus}, {Komin}, {Kosack}, {Krakau}, {Krayzel}, {Kr{\"u}ger}, {Laffon},
  {Lamanna}, {Lefaucheur}, {Lefranc}, {Lemi{\`e}re}, {Lemoine-Goumard},
  {Lenain}, {Lohse}, {Lopatin}, {Lu}, {Marandon}, {Marcowith}, {Marx},
  {Maurin}, {Maxted}, {Mayer}, {McComb}, {M{\'e}hault}, {Meintjes}, {Menzler},
  {Meyer}, {Mitchell}, {Moderski}, {Mohamed}, {Mor{\r{a}}}, {Moulin}, {Murach},
  {de Naurois}, {Niemiec}, {Nolan}, {Oakes}, {Odaka}, {Ohm}, {Opitz},
  {Ostrowski}, {Oya}, {Panter}, {Parsons}, {Arribas}, {Pekeur}, {Pelletier},
  {Petrucci}, {Peyaud}, {Pita}, {Poon}, {P{\"u}hlhofer}, {Punch},
  {Quirrenbach}, {Raab}, {Reichardt}, {Reimer}, {Reimer}, {Renaud}, {de los
  Reyes}, {Rieger}, {Romoli}, {Rosier-Lees}, {Rowell}, {Rudak}, {Rulten},
  {Sahakian}, {Salek}, {Sanchez}, {Santangelo}, {Schlickeiser},
  {Sch{\"u}ssler}, {Schulz}, {Schwanke}, {Schwarzburg}, {Schwemmer}, {Sol},
  {Spanier}, {Spengler}, {Spies}, {Stawarz}, {Steenkamp}, {Stegmann},
  {Stinzing}, {Stycz}, {Sushch}, {Tavernet}, {Tavernier}, {Taylor}, {Terrier},
  {Tluczykont}, {Trichard}, {Valerius}, {van Eldik}, {van Soelen},
  {Vasileiadis}, {Veh}, {Venter}, {Viana}, {Vincent}, {Vink}, {V{\"o}lk},
  {Volpe}, {Vorster}, {Vuillaume}, {Wagner}, {Wagner}, {Wagner}, {Ward},
  {Weidinger}, {Weitzel}, {White}, {Wierzcholska}, {Willmann}, {W{\"o}rnlein},
  {Wouters}, {Yang}, {Zabalza}, {Zaborov}, {Zacharias}, {Zdziarski}, {Zech}, \&
  {Zechlin}}]{2015A&A...574A.100H}
{H.~E.~S.~S. Collaboration}, {Abramowski}, A., {Aharonian}, F., {et~al.} 2015,
  \aap, 574, A100, \dodoi{10.1051/0004-6361/201425070}

\bibitem[{{H.~E.~S.~S. Collaboration} {et~al.}(2018){H.~E.~S.~S.
  Collaboration}, {Abdalla}, {Abramowski}, {Aharonian}, {Ait Benkhali},
  {Ang{\"u}ner}, {Arakawa}, {Arrieta}, {Aubert}, {Backes}, {Balzer}, {Barnard},
  {Becherini}, {Becker Tjus}, {Berge}, {Bernhard}, {Bernl{\"o}hr}, {Blackwell},
  {B{\"o}ttcher}, {Boisson}, {Bolmont}, {Bonnefoy}, {Bordas}, {Bregeon},
  {Brun}, {Brun}, {Bryan}, {B{\"u}chele}, {Bulik}, {Capasso}, {Carrigan},
  {Caroff}, {Carosi}, {Casanova}, {Cerruti}, {Chakraborty}, {Chaves}, {Chen},
  {Chevalier}, {Colafrancesco}, {Condon}, {Conrad}, {Davids}, {Decock}, {Deil},
  {Devin}, {deWilt}, {Dirson}, {Djannati-Ata{\"\i}}, {Domainko}, {Donath},
  {Drury}, {Dutson}, {Dyks}, {Edwards}, {Egberts}, {Eger}, {Emery},
  {Ernenwein}, {Eschbach}, {Farnier}, {Fegan}, {Fernandes}, {Fiasson},
  {Fontaine}, {F{\"o}rster}, {Funk}, {F{\"u}{\ss}ling}, {Gabici}, {Gallant},
  {Garrigoux}, {Gast}, {Gat{\'e}}, {Giavitto}, {Giebels}, {Glawion},
  {Glicenstein}, {Gottschall}, {Grondin}, {Hahn}, {Haupt}, {Hawkes},
  {Heinzelmann}, {Henri}, {Hermann}, {Hinton}, {Hofmann}, {Hoischen}, {Holch},
  {Holler}, {Horns}, {Ivascenko}, {Iwasaki}, {Jacholkowska}, {Jamrozy},
  {Jankowsky}, {Jankowsky}, {Jingo}, {Jouvin}, {Jung-Richardt}, {Kastendieck},
  {Katarzy{\'n}ski}, {Katsuragawa}, {Katz}, {Kerszberg}, {Khangulyan},
  {Kh{\'e}lifi}, {King}, {Klepser}, {Klochkov}, {Klu{\'z}niak}, {Komin},
  {Kosack}, {Krakau}, {Kraus}, {Kr{\"u}ger}, {Laffon}, {Lamanna}, {Lau},
  {Lees}, {Lefaucheur}, {Lemi{\`e}re}, {Lemoine-Goumard}, {Lenain}, {Leser},
  {Lohse}, {Lorentz}, {Liu}, {L{\'o}pez-Coto}, {Lypova}, {Marandon},
  {Malyshev}, {Marcowith}, {Mariaud}, {Marx}, {Maurin}, {Maxted}, {Mayer},
  {Meintjes}, {Meyer}, {Mitchell}, {Moderski}, {Mohamed}, {Mohrmann},
  {Mor{\r{a}}}, {Moulin}, {Murach}, {Nakashima}, {de Naurois}, {Ndiyavala},
  {Niederwanger}, {Niemiec}, {Oakes}, {O'Brien}, {Odaka}, {Ohm}, {Ostrowski},
  {Oya}, {Padovani}, {Panter}, {Parsons}, {Paz Arribas}, {Pekeur}, {Pelletier},
  {Perennes}, {Petrucci}, {Peyaud}, {Piel}, {Pita}, {Poireau}, {Poon},
  {Prokhorov}, {Prokoph}, {P{\"u}hlhofer}, {Punch}, {Quirrenbach}, {Raab},
  {Rauth}, {Reimer}, {Reimer}, {Renaud}, {de los Reyes}, {Rieger}, {Rinchiuso},
  {Romoli}, {Rowell}, {Rudak}, {Rulten}, {Safi-Harb}, {Sahakian}, {Saito},
  {Sanchez}, {Santangelo}, {Sasaki}, {Schandri}, {Schlickeiser},
  {Sch{\"u}ssler}, {Schulz}, {Schwanke}, {Schwemmer}, {Seglar-Arroyo},
  {Settimo}, {Seyffert}, {Shafi}, {Shilon}, {Shiningayamwe}, {Simoni}, {Sol},
  {Spanier}, {Spir-Jacob}, {Stawarz}, {Steenkamp}, {Stegmann}, {Steppa},
  {Sushch}, {Takahashi}, {Tavernet}, {Tavernier}, {Taylor}, {Terrier},
  {Tibaldo}, {Tiziani}, {Tluczykont}, {Trichard}, {Tsirou}, {Tsuji}, {Tuffs},
  {Uchiyama}, {van der Walt}, {van Eldik}, {van Rensburg}, {van Soelen},
  {Vasileiadis}, {Veh}, {Venter}, {Viana}, {Vincent}, {Vink}, {Voisin},
  {V{\"o}lk}, {Vuillaume}, {Wadiasingh}, {Wagner}, {Wagner}, {Wagner}, {White},
  {Wierzcholska}, {Willmann}, {W{\"o}rnlein}, {Wouters}, {Yang}, {Zaborov},
  {Zacharias}, {Zanin}, {Zdziarski}, {Zech}, {Zefi}, {Ziegler}, {Zorn}, \&
  {{\.Z}ywucka}}]{2018A&A...612A...1H}
{H.~E.~S.~S. Collaboration}, {Abdalla}, H., {Abramowski}, A., {et~al.} 2018,
  \aap, 612, A1, \dodoi{10.1051/0004-6361/201732098}

\bibitem[{{Jackson} {et~al.}(2006){Jackson}, {Rathborne}, {Shah}, {Simon},
  {Bania}, {Clemens}, {Chambers}, {Johnson}, {Dormody}, {Lavoie}, \&
  {Heyer}}]{2006ApJS..163..145J}
{Jackson}, J.~M., {Rathborne}, J.~M., {Shah}, R.~Y., {et~al.} 2006, \apjs, 163,
  145, \dodoi{10.1086/500091}

\bibitem[{{Kafexhiu} {et~al.}(2014){Kafexhiu}, {Aharonian}, {Taylor}, \&
  {Vila}}]{2014PhRvD..90l3014K}
{Kafexhiu}, E., {Aharonian}, F., {Taylor}, A.~M., \& {Vila}, G.~S. 2014, \prd,
  90, 123014, \dodoi{10.1103/PhysRevD.90.123014}

\bibitem[{{Kaur} {et~al.}(2019){Kaur}, {Falcone}, {Stroh}, {Kennea}, \&
  {Ferrara}}]{2019ApJ...887...18K}
{Kaur}, A., {Falcone}, A.~D., {Stroh}, M.~D., {Kennea}, J.~A., \& {Ferrara},
  E.~C. 2019, \apj, 887, 18, \dodoi{10.3847/1538-4357/ab4ceb}

\bibitem[{{Lande} {et~al.}(2012){Lande}, {Ackermann}, {Allafort}, {Ballet},
  {Bechtol}, {Burnett}, {Cohen-Tanugi}, {Drlica-Wagner}, {Funk}, {Giordano},
  {Grondin}, {Kerr}, \& {Lemoine-Goumard}}]{2012ApJ...756....5L}
{Lande}, J., {Ackermann}, M., {Allafort}, A., {et~al.} 2012, \apj, 756, 5,
  \dodoi{10.1088/0004-637X/756/1/5}

\bibitem[{{Lau} {et~al.}(2017){Lau}, {Rowell}, {Burton}, {Fukui}, {Aharonian},
  {Oya}, {Vink}, {Ohm}, \& {Casanova}}]{2017MNRAS.464.3757L}
{Lau}, J.~C., {Rowell}, G., {Burton}, M.~G., {et~al.} 2017, \mnras, 464, 3757,
  \dodoi{10.1093/mnras/stw2692}

\bibitem[{{Liu} {et~al.}(2019){Liu}, {Anderson}, {McIntyre}, {Anish Roshi},
  {Churchwell}, {Minchin}, \& {Terzian}}]{2019ApJS..240...14L}
{Liu}, B., {Anderson}, L.~D., {McIntyre}, T., {et~al.} 2019, \apjs, 240, 14,
  \dodoi{10.3847/1538-4365/aaef8e}

\bibitem[{{Ma{\'\i}z Apell{\'a}niz} {et~al.}(2016){Ma{\'\i}z Apell{\'a}niz},
  {Sota}, {Arias}, {Barb{\'a}}, {Walborn}, {Sim{\'o}n-D{\'\i}az}, {Negueruela},
  {Marco}, {Le{\~a}o}, {Herrero}, {Gamen}, \& {Alfaro}}]{2016ApJS..224....4M}
{Ma{\'\i}z Apell{\'a}niz}, J., {Sota}, A., {Arias}, J.~I., {et~al.} 2016,
  \apjs, 224, 4, \dodoi{10.3847/0067-0049/224/1/4}

\bibitem[{{Manchester} {et~al.}(2005){Manchester}, {Hobbs}, {Teoh}, \&
  {Hobbs}}]{2005AJ....129.1993M}
{Manchester}, R.~N., {Hobbs}, G.~B., {Teoh}, A., \& {Hobbs}, M. 2005, \aj, 129,
  1993, \dodoi{10.1086/428488}

\bibitem[{{Mattox} {et~al.}(1996){Mattox}, {Bertsch}, {Chiang}, {Dingus},
  {Digel}, {Esposito}, {Fierro}, {Hartman}, {Hunter}, {Kanbach}, {Kniffen},
  {Lin}, {Macomb}, {Mayer-Hasselwander}, {Michelson}, {von Montigny},
  {Mukherjee}, {Nolan}, {Ramanamurthy}, {Schneid}, {Sreekumar}, {Thompson}, \&
  {Willis}}]{1996ApJ...461..396M}
{Mattox}, J.~R., {Bertsch}, D.~L., {Chiang}, J., {et~al.} 1996, \apj, 461, 396,
  \dodoi{10.1086/177068}

\bibitem[{{Planck Collaboration} {et~al.}(2016){Planck Collaboration},
  {Arnaud}, {Ashdown}, {Atrio-Barandela}, {Aumont}, {Baccigalupi}, {Banday},
  {Barreiro}, {Battaner}, {Benabed}, {Benoit-L{\'e}vy}, {Bernard},
  {Bersanelli}, {Bielewicz}, {Bobin}, {Bond}, {Borrill}, {Bouchet}, {Brogan},
  {Burigana}, {Cardoso}, {Catalano}, {Chamballu}, {Chiang}, {Christensen},
  {Colombi}, {Colombo}, {Crill}, {Curto}, {Cuttaia}, {Davies}, {Davis}, {de
  Bernardis}, {de Rosa}, {de Zotti}, {Delabrouille}, {D{\'e}sert}, {Dickinson},
  {Diego}, {Donzelli}, {Dor{\'e}}, {Dupac}, {En{\ss}lin}, {Eriksen}, {Finelli},
  {Forni}, {Frailis}, {Fraisse}, {Franceschi}, {Galeotta}, {Ganga}, {Giard},
  {Giraud-H{\'e}raud}, {Gonz{\'a}lez-Nuevo}, {G{\'o}rski}, {Gregorio},
  {Gruppuso}, {Hansen}, {Harrison}, {Hern{\'a}ndez-Monteagudo}, {Herranz},
  {Hildebrandt}, {Hobson}, {Holmes}, {Huffenberger}, {Jaffe}, {Jaffe},
  {Keih{\"a}nen}, {Keskitalo}, {Kisner}, {Kneissl}, {Knoche}, {Kunz},
  {Kurki-Suonio}, {L{\"a}hteenm{\"a}ki}, {Lamarre}, {Lasenby}, {Lawrence},
  {Leonardi}, {Liguori}, {Lilje}, {Linden-V{\o}rnle}, {L{\'o}pez-Caniego},
  {Lubin}, {Maino}, {Maris}, {Marshall}, {Martin},
  {Mart{\'\i}nez-Gonz{\'a}lez}, {Masi}, {Matarrese}, {Mazzotta}, {Melchiorri},
  {Mendes}, {Mennella}, {Migliaccio}, {Miville-Desch{\^e}nes}, {Moneti},
  {Montier}, {Morgante}, {Mortlock}, {Munshi}, {Murphy}, {Naselsky}, {Nati},
  {Noviello}, {Novikov}, {Novikov}, {Oppermann}, {Oxborrow}, {Pagano}, {Pajot},
  {Paladini}, {Pasian}, {Peel}, {Perdereau}, {Perrotta}, {Piacentini}, {Piat},
  {Pietrobon}, {Plaszczynski}, {Pointecouteau}, {Polenta}, {Popa}, {Pratt},
  {Puget}, {Rachen}, {Reach}, {Reich}, {Reinecke}, {Remazeilles}, {Renault},
  {Rho}, {Ricciardi}, {Riller}, {Ristorcelli}, {Rocha}, {Rosset}, {Roudier},
  {Rusholme}, {Sandri}, {Savini}, {Scott}, {Stolyarov}, {Sutton}, {Suur-Uski},
  {Sygnet}, {Tauber}, {Terenzi}, {Toffolatti}, {Tomasi}, {Tristram}, {Tucci},
  {Umana}, {Valenziano}, {Valiviita}, {Van Tent}, {Vielva}, {Villa}, {Wade},
  {Yvon}, {Zacchei}, \& {Zonca}}]{2016A&A...586A.134P}
{Planck Collaboration}, {Arnaud}, M., {Ashdown}, M., {et~al.} 2016, \aap, 586,
  A134, \dodoi{10.1051/0004-6361/201425022}

\bibitem[{{Principe} {et~al.}(2020){Principe}, {Mitchell}, {Caroff}, {Hinton},
  {Parsons}, \& {Funk}}]{2020A&A...640A..76P}
{Principe}, G., {Mitchell}, A.~M.~W., {Caroff}, S., {et~al.} 2020, \aap, 640,
  A76, \dodoi{10.1051/0004-6361/202038375}

\bibitem[{{Reid} {et~al.}(2019){Reid}, {Menten}, {Brunthaler}, {Zheng}, {Dame},
  {Xu}, {Li}, {Sakai}, {Wu}, {Immer}, {Zhang}, {Sanna}, {Moscadelli}, {Rygl},
  {Bartkiewicz}, {Hu}, {Quiroga-Nu{\~n}ez}, \& {van
  Langevelde}}]{2019ApJ...885..131R}
{Reid}, M.~J., {Menten}, K.~M., {Brunthaler}, A., {et~al.} 2019, \apj, 885,
  131, \dodoi{10.3847/1538-4357/ab4a11}

\bibitem[{{Shibata} {et~al.}(2011){Shibata}, {Ishikawa}, \&
  {Sekiguchi}}]{2011ApJ...727...38S}
{Shibata}, T., {Ishikawa}, T., \& {Sekiguchi}, S. 2011, \apj, 727, 38,
  \dodoi{10.1088/0004-637X/727/1/38}

\bibitem[{{Simon} {et~al.}(2001){Simon}, {Jackson}, {Clemens}, {Bania}, \&
  {Heyer}}]{2001ApJ...551..747S}
{Simon}, R., {Jackson}, J.~M., {Clemens}, D.~P., {Bania}, T.~M., \& {Heyer},
  M.~H. 2001, \apj, 551, 747, \dodoi{10.1086/320230}

\bibitem[{{Stroh} \& {Falcone}(2013)}]{2013ApJS..207...28S}
{Stroh}, M.~C., \& {Falcone}, A.~D. 2013, \apjs, 207, 28,
  \dodoi{10.1088/0067-0049/207/2/28}

\bibitem[{{Sun} {et~al.}(2020){Sun}, {Yang}, {Liang}, {Peng}, {Zhang}, {Wang},
  \& {Aharonian}}]{2020A&A...639A..80S}
{Sun}, X.-N., {Yang}, R.-Z., {Liang}, Y.-F., {et~al.} 2020, \aap, 639, A80,
  \dodoi{10.1051/0004-6361/202037580}

\bibitem[{{Supan} {et~al.}(2018){Supan}, {Castelletti}, {Peters}, \&
  {Kassim}}]{2018A&A...616A..98S}
{Supan}, L., {Castelletti}, G., {Peters}, W.~M., \& {Kassim}, N.~E. 2018, \aap,
  616, A98, \dodoi{10.1051/0004-6361/201832995}

\bibitem[{{van den Bergh} \& {McClure}(1994)}]{1994ApJ...425..205V}
{van den Bergh}, S., \& {McClure}, R.~D. 1994, \apj, 425, 205,
  \dodoi{10.1086/173975}

\bibitem[{{Wang} {et~al.}(2018){Wang}, {Bihr}, {Rugel}, {Beuther}, {Johnston},
  {Ott}, {Soler}, {Brunthaler}, {Anderson}, {Urquhart}, {Klessen}, {Linz},
  {McClure-Griffiths}, {Glover}, {Menten}, {Bigiel}, {Hoare}, \&
  {Longmore}}]{2018A&A...619A.124W}
{Wang}, Y., {Bihr}, S., {Rugel}, M., {et~al.} 2018, \aap, 619, A124,
  \dodoi{10.1051/0004-6361/201833642}

\bibitem[{{Wright} {et~al.}(2010){Wright}, {Eisenhardt}, {Mainzer}, {Ressler},
  {Cutri}, {Jarrett}, {Kirkpatrick}, {Padgett}, {McMillan}, {Skrutskie},
  {Stanford}, {Cohen}, {Walker}, {Mather}, {Leisawitz}, {Gautier}, {McLean},
  {Benford}, {Lonsdale}, {Blain}, {Mendez}, {Irace}, {Duval}, {Liu}, {Royer},
  {Heinrichsen}, {Howard}, {Shannon}, {Kendall}, {Walsh}, {Larsen}, {Cardon},
  {Schick}, {Schwalm}, {Abid}, {Fabinsky}, {Naes}, \&
  {Tsai}}]{2010AJ....140.1868W}
{Wright}, E.~L., {Eisenhardt}, P. R.~M., {Mainzer}, A.~K., {et~al.} 2010, \aj,
  140, 1868, \dodoi{10.1088/0004-6256/140/6/1868}

\bibitem[{{Xing} {et~al.}(2019){Xing}, {Wang}, {Zhang}, \&
  {Chen}}]{2019PASJ...71...77X}
{Xing}, Y., {Wang}, Z., {Zhang}, X., \& {Chen}, Y. 2019, \pasj, 71, 77,
  \dodoi{10.1093/pasj/psz056}

\bibitem[{{Yuan} {et~al.}(2013){Yuan}, {Funk}, {J{\'o}hannesson}, {Lande},
  {Tibaldo}, \& {Uchiyama}}]{2013ApJ...779..117Y}
{Yuan}, Y., {Funk}, S., {J{\'o}hannesson}, G., {et~al.} 2013, \apj, 779, 117,
  \dodoi{10.1088/0004-637X/779/2/117}

\bibitem[{{Zabalza}(2015)}]{naima}
{Zabalza}, V. 2015, Proceedings of The 34th International Cosmic Ray
  Conference, The Hague, The Netherlands, 236, 922, \dodoi{10.22323/1.236.0922}

\end{thebibliography}

\end{document}